\documentclass[11pt, a4paper]{article}
\pdfoutput=1

\usepackage[text={6in, 9in}, centering]{geometry}
\usepackage[tbtags]{amsmath} % tbtags places tags in split environments on the last line
\usepackage{amsfonts,amssymb,amsthm,amscd}
\usepackage[utf8]{inputenc}
\usepackage[T1]{fontenc}
\usepackage{graphicx}
\usepackage{lmodern}
\usepackage{microtype}
\usepackage[pdfborder={0 0 0}]{hyperref}
\usepackage{color}
\usepackage{dsfont} % for \mathds{1}
\usepackage{marginnote} % for \marginnote, a replacement for \marginpar that works everywhere
\usepackage{mathtools} % for \smashoperator
\usepackage{ifthen} % for \ifthenelse
\usepackage{tensor}
\usepackage{paralist} % for compactitem, etc.

\numberwithin{equation}{section}
\DeclareMathOperator{\Ad}{Ad}
\DeclareMathOperator{\ad}{ad}

\DeclareMathOperator{\tr}{Tr}

\newcommand{\inv}[0]{{-1}}

\newcommand{\cif}[0]{\mathcal{C}^\infty}
\newcommand{\surf}[0]{S_{g,n}}
\newcommand{\oo}[0]{\otimes}
\newcommand{\ee}[0]{\varepsilon}

\newcommand{\boldify}[1]{\ensuremath{\boldsymbol{#1}}}
\def\ba{\boldify{a}}

\def\be{\boldify{e}}
\def\bj{\boldify{j}}
\def\bk{\boldify{k}}
\def\bm{\boldify{m}}

\def\bp{\boldify{p}}
\def\bq{\boldify{q}}
\def\br{\boldify{r}}

\def\bt{\boldify{t}}
\def\bu{\boldify{u}}
\def\bv{\boldify{v}}
\def\bw{\boldify{w}}
\def\bx{\boldify{x}}
\def\by{\boldify{y}}

\newcommand{\hattify}[1]{\ensuremath{\hat{#1}}}
\def\hp{\hattify{p}}

\def\hbp{\hattify{\boldify{p}}}

\newcommand{\bpres}{\bp_R}

\newcommand{\bjres}{\bj_R}

\newcommand{\RR}{\mathbb{R}}
\newcommand{\CC}{\mathbb{C}}

\newcommand{\ki}[0]{{K_i}}
\newcommand{\mi}[0]{{M_i}}
\newcommand{\ai}[0]{{A_i}}
\newcommand{\bi}[0]{{B_i}}

\newcommand{\phasespace}{\ensuremath{\mathcal{P}_{\mathrm{ext}}}}
\newcommand{\csurface}{\ensuremath{\Sigma}}
\newcommand{\gaugeinvpspace}{\ensuremath{\mathcal{P}}}
\newcommand{\gaugetrafos}{\mathcal{G}}

\newcommand{\issue}[2][]{%
  \marginnote{\footnotesize\raggedright%
    \textbf{\textcolor{red}{issue}}%
    \ifthenelse{\equal{#1}{}}{}{: #1}}%
  \textcolor{red}{#2}}

\newcommand{\defeq}{\coloneqq}
\newcommand{\diffd}{\mathrm{d}}

\newcommand{\tdiff}[2]{\frac{\diffd #1}{\diffd #2}}
\newcommand{\tdiffat}[3]{\left.\tdiff{#1}{#2}\right|_{#3}}

\newcommand{\idad}[1]{\bigl(\mathds{1}-\Ad(u_{#1})\bigr)}

\newcommand{\tidad}[2]{\tensor{\idad{#1}}{#2}}
\newcommand{\idadi}[1]{\bigl(\mathds{1}-\Ad(u_{#1}^\inv)\bigr)}

\newcommand{\tidadi}[2]{\tensor{\idadi{#1}}{#2}}

\newtheorem{theorem}{Theorem}[section]

\newtheorem{corollary}[theorem]{Corollary}

\newcommand{\allowlinebreakhere}{\linebreak[0]}
\newcommand{\allowpagebreakhere}{\displaybreak[0]}

\newcommand{\email}[1]{\href{mailto:#1}{\nolinkurl{#1}}}

\newcommand{\cf}{cf.\ }

\newcommand{\ie}{i.e.\ }

\begin{document}

%
% title
%
\begin{flushright}
  ZMP-HH/10-22 \\
  Hamburger Beiträge zur Mathematik Nr. 393
\end{flushright}

\begin{center}
  {\LARGE\sc
   Gauge fixing in (2+1)-gravity: \\[0.1em] Dirac bracket and spacetime geometry}

  \vspace{2em}

  {\large
   C.~Meusburger\footnote{\email{catherine.meusburger@uni-hamburg.de}} \qquad\qquad
   T.~Sch\"onfeld\footnote{\email{torsten.schoenfeld@uni-hamburg.de}}}

  \vspace{1em}

  Fachbereich Mathematik, \\ Universit\"at Hamburg \\
  Bundesstra\ss e 55 \\ D-20146 Hamburg, Germany

  \vspace{1em}

  December 8 2010

  \vspace{2em}

  \begin{abstract}
    We consider (2+1)-gravity with vanishing cosmological constant as a
    constrained dynamical system.  By applying Dirac's gauge fixing procedure, we
    implement the constraints  and determine the Dirac bracket on
    the gauge-invariant phase space. The chosen gauge fixing conditions have a
    natural physical interpretation and  specify an observer in
    the spacetime. We derive explicit expressions for the resulting Dirac
    brackets and discuss their geometrical interpretation.  In particular, we
    show that specifying an observer with respect to two point particles gives
    rise to conical spacetimes, whose deficit angle and time shift are
    determined, respectively, by the relative velocity and minimal distance of
    the two particles.
  \end{abstract}
\end{center}

%%%%%%%%%%%%%%%%%%%%%%%%%%%%%%%%%%%

\section{Introduction}
\label{sec:intro}

The diffeomorphism symmetry of general relativity has profound consequences for its quantisation. Its physical implications, such as the equivalence of observers and the absence of a fixed spacetime background, lead to conceptual problems in the construction of the quantum theory. In the description of its phase space, the diffeomorphism symmetry manifests itself through the presence of constraints. The implementation of these constraints in a quantum theory of gravity remains one of the fundamental problems of quantum gravity.

As a consequence, investigations of  the structure of ``quantum spacetimes'' tend to be based on either phenomenological  models or on a partially quantised version of the theory in which the constraints are not fully implemented. The former  have the draw-back  that their relation to the theory of general relativity  is not immediately apparent. For the latter it remains unclear whether results and conclusions that are drawn from an extended, not fully diffeomorphism-invariant version of the theory
 remain valid after the full implementation of the constraints. Although there are recent  results on  constraint implementation and gauge fixing in Regge-Calculus \cite{Dittrich:2008ar,Bahr:2009ku,Dittrich:2010fj,Dittrich:2010ey} and in state sum models of gravity (``spin foams'')  \cite{Buffenoir:2004vx}, a full implementation of the constraints in  (3+1)-dimensional quantum gravity seems currently not within reach.

In (2+1)-dimensions, the theory simplifies considerably, while the conceptual issues and the problems related to constraint implementation are still present. Gravity in (2+1)-dimensions  therefore serves as a toy model that allows one to investigate these issues  in a fully quantised version of the theory. Important progress in the quantisation of the theory was achieved with quantisation formalisms based on the representation theory of quantum groups such as the combinatorial quantisation formalism \cite{Alekseev:1995ab,Alekseev:1996aa,Buffenoir:1995aa}, the Turaev-Viro model \cite{Turaev:1992hq,Barrett:1996aa} and the Reshetikhin-Turaev-Witten invariants \cite{Witten:1989aa,Reshetikhin:1991aa} arising in the Chern-Simons formulation of the theory.

As these formalisms achieve constraint implementation via the representation theory of quantum groups, they enjoyed great success in the applications where the relevant quantum groups are compact. However, except for Euclidean (2+1)-gravity with positive cosmological constant,  the quantum groups arising in (2+1)-gravity are non-compact. The resulting difficulties in the combinatorial quantisation formalism  were resolved for Lorentzian gravity with positive cosmological constant \cite{Buffenoir:2002aa} and partially for the case of vanishing cosmological constant \cite{Meusburger:2008aa,Meusburger:2010bc}. 
However,  the  representation-theoretical complications that arise in these cases suggest that it could be advantageous to
first implement the constraints in the classical theory via gauge fixing and quantise the theory afterwards. For the cases in which a ``constraint implementation after quantisation'' approach is possible, this is also of interest, as it would allow one to compare the two approaches and determine if the resulting quantum theories are equivalent.

A further motivation to investigate gauge fixing  is an improved understanding of the interplay between gauge fixing and quantum group symmetry in (2+1)-gravity. 
With the exception of \cite{Buffenoir:2005zi}, which investigates quantum group symmetries in a gauge-invariant quantised $SL(2,\mathbb C)$-Chern-Simons theory with punctures,
most investigations of quantum group symmetries
 in quantum gravity  focus on extended, not fully diffeomorphism-invariant versions of the quantum theory. It
therefore remains unclear if these quantum group symmetries are compatible with diffeomorphism invariance and survive the implementation of constraints.
As the presence of  quantum group symmetries manifests itself through the associated Poisson-Lie symmetries in the classical theory, constructing the Dirac bracket for (2+1)-gravity should   allow one to draw conclusions  about the presence of quantum group symmetries in a fully gauge-invariant quantum theory.

Independently of its role in quantisation, gauge fixing in classical (2+1)-gravity 
is of interest with respect to its physical implications. In particular, there are strong indications that it is directly related to the inclusion of observers into the theory \cite{Meusburger:2009aa}, which is a fundamental problem in quantum gravity. Constructing the Dirac bracket should therefore lead to a formulation of the classical theory  which includes observers. This would be helpful for its physical interpretation  and could be used to include observers in the associated quantum theory

In this article, we apply Dirac's gauge fixing procedure to the Chern-Simons formulation of Lorentzian (2+1)-gravity with vanishing cosmological constant. We consider  spacetimes of topology $\RR\times S_{g,n}$, where $S_{g,n}$ is a compact oriented surface of genus $g$  with $n$ punctures that represent massive point particles with spin\footnote{The spacetimes are also subject to certain geometrical restrictions that are detailed in Section \ref{sec:gfix}}. The gauge-invariant (reduced) phase space of these models is  a moduli space of flat connections on the spatial surface $S_{g,n}$.
The Poisson structure on the moduli space of flat connections can be described in terms of a Poisson structure on an enlarged ambient space
 from which it is obtained by imposing six first-class constraints \cite{Fock:1998aa,Meusburger:2003aa}.
 The parameters used in this description have a direct physical interpretation and can be related to measurements  by observers in a spacetime \cite{Meusburger:2009aa}.

 We determine the Dirac bracket associated to this Poisson structure and construct the gauge-invariant (reduced) phase space.   Our starting point is the observation that the gauge transformations generated by these six first-class constraints correspond to transformations that relate different inertial observers  in the spacetime. Eliminating this gauge freedom via gauge fixing conditions thus amounts to fixing an observer in the spacetime.
 We consider two different types of gauge fixing conditions. The first specifies an observer with respect to the motion of two particles in the spacetime. The second characterises an observer with respect to the geometry of a handle of the surface $\surf$.

 We show that the resulting Dirac bracket has a direct physical interpretation in terms of spacetime geometry. For the first gauge fixing condition, it leads to a spacetime which is effectively conical. The opening angle and the time shift of the associated cone are given, respectively, by the relative velocity and the minimal distance of the particles.  For the second gauge fixing condition,  we obtain an effectively Minkowskian spacetime whose global degrees of freedom are determined by the geometry of the handles.

We then give a detailed analysis of the Dirac bracket and its physical interpretation. In particular, we show that the Dirac bracket  can be obtained from the non-gauge-fixed Poisson bracket for a spacetime with, respectively,  $n-2$ particles or $g-1$ handles via a global translation. For the gauge fixing conditions based on point particles, this allows one to continue the bracket beyond the constraint surface.

Our paper is structured as follows. In Section \ref{sec:background} we give a brief introduction to the geometry of (2+1)-dimensional spacetimes and their description in terms of a Chern-Simons gauge theory. We summarise the relevant results for their phase space and Poisson structure  and introduce Fock and Rosly's  Poisson structure  \cite{Fock:1998aa}, which describes the Poisson structure
on the moduli space of flat connections in terms of an enlarged ambient space.

In Section \ref{sec:gauge} we investigate
this Poisson structure  from the viewpoint of constrained systems. We discuss the constraints arising in this description and  the associated gauge transformations and show that the latter
can be interpreted as transformations that relate different inertial observers in the spacetime.
In Section \ref{sec:gfix}, we motivate and define our gauge fixing conditions and show that they correspond to specifying an observer in the spacetime.

Sections \ref{sec:results} and \ref{sec:interpretation} contain the main results of our paper. In Section \ref{sec:results}, we implement the constraints and gauge fixing conditions and derive explicit expressions for the associated Dirac bracket.
Section \ref{sec:interpretation} gives a detailed analysis of the resulting Poisson structure  and its implications in terms of spacetime geometry. We show that the gauge fixing based on two point particles
 leads to an effectively conical spacetime and the gauge fixing  based on handles to a spacetime that is effectively  Minkowskian.
  For both cases, we show how the Dirac bracket can be obtained from the original Poisson bracket for a reduced system  via a global translation.
Section \ref{sec:outlook} contains our outlook and conclusions.

%%%%%%%%%%%%%%%%%%%%%%%%%%%%%%%%%%%

\section{Gravity in (2+1) dimensions}
\label{sec:background}

\subsection{Notation and conventions}
\label{sec:notation}

Throughout the paper, we employ Einstein's summation convention. Unless stated
otherwise, all indices run from 0 to 2 and are raised and lowered with the
Minkowski metric $\eta=\text{diag}(1,-1,-1)$. We denote by $\ee_{abc}$ the
totally antisymmetric tensor in three dimensions with the convention
$\ee_{012}=1$.  For three-vectors $\bx,\by\in\RR^3$, we use the notation
$\bx\cdot\by=\eta_{ab}x^ay^b$, and we denote by $\bx\wedge\by$ the three-vector
with components $(\bx\wedge\by)^a=\ee^{abc}x_by_c$.

The proper orthochronous Lorentz group in three dimensions is the group $SO(2,1)^+\cong PSL(2,\RR)$.
We fix a set of generators $J_a$, $a=0,1,2$, of its Lie algebra
$\mathfrak{so}(2,1)\cong\mathfrak{sl}(2,\RR)$ such that the Lie bracket takes
the form
\begin{equation}\label{lorbrac}
  [J_a,J_b]=\tensor{\ee}{_a_b^c}J_c.
\end{equation}
In the representation by $\mathfrak{sl}(2,\RR)$ matrices, such a set of
generators is given by
\begin{equation}\label{jmatrix}
  J_0=\tfrac 1 2 \begin{pmatrix}0 & 1 \\ -1 & 0\end{pmatrix},\qquad
  J_1=\tfrac 1 2 \begin{pmatrix}1 & 0\\ 0 & -1\end{pmatrix},\qquad
  J_2=\tfrac 1 2 \begin{pmatrix}0 & 1\\ 1 & 0\end{pmatrix}.
\end{equation}
As the generators satisfy the relation $J_a\cdot J_b=-\tfrac 1 4 \eta_{ab}
\mathds{1}+\tfrac 1 2 \tensor{\ee}{_a_b^c}J_c$,
%\begin{equation}\label{multsl}
%  J_a\cdot J_b=-\tfrac 1 4 \eta_{ab} \mathds{1}+\tfrac 1 2 \tensor{\ee}{_a_b^c}J_c,
%\end{equation}
the exponential map $\exp: \mathfrak{sl}(2,\mathbb R)\rightarrow SL(2,\RR)$
takes the form
\begin{equation}\label{expsl}
  \exp(p^a J_a)=
    \begin{cases}
      \cos\tfrac \mu 2 \, \mathds{1}+2\sin\tfrac \mu 2 \, \hp^aJ_a & \text{if $\bp^2>0$},\\
      \cosh\tfrac \mu 2 \, \mathds{1}+2\sinh\tfrac \mu 2 \, \hp^aJ_a & \text{if $\bp^2<0$},\\
      \mathds{1}+p^aJ_a & \text{if $\bp^2=0$},
    \end{cases}
\end{equation}
where
\begin{equation}
  \mu\defeq\sqrt{\bigl|\bp^2\bigr|},
  \qquad
  \hbp\defeq \begin{cases}
    \bp/\mu & \text{if $\bp^2 \ne 0$}, \\
    \bp     & \text{if $\bp^2 = 0$}.
  \end{cases}
\end{equation}
Introducing the notation
\begin{align}\label{gencossin}
  c_p(\mu)\defeq\begin{cases}
    \cos\mu  & \text{if $\bp^2>0$},\\
    \cosh\mu & \text{if $\bp^2<0$},\\
    1        & \text{if $\bp^2=0$},
  \end{cases}\qquad
  s_p(\mu)\defeq\begin{cases}
    \sin\mu  & \text{if $\bp^2>0$},\\
    \sinh\mu & \text{if $\bp^2<0$},\\
    0        & \text{if $\bp^2=0$},
  \end{cases}
\end{align}
we can rewrite \eqref{expsl} as
\begin{align}
  \exp(p^aJ_a)=c_p(\tfrac{\mu}{2})\mathds{1} + \frac{s_p({\mu}/{2})}{\mu/2} p^a J_a,
\end{align}
where the expression for lightlike vectors $\bp$ is given by the
limit $ \lim_{\bp^2 \to 0} \frac{s_p({\mu}/{2})}{\mu/2} =1$.
%\begin{equation}
%  \lim_{\bp^2 \to 0} \frac{s_p({\mu}/{2})}{\mu/2} =1.
%\end{equation}

Note that the exponential map $\exp: \mathfrak{sl}(2,\mathbb R)\rightarrow
SL(2,\RR)$ is neither surjective nor injective. The exponential map
$\exp:\mathfrak{sl}(2,\RR)\rightarrow PSL(2,\RR)$ is surjective, but not
injective.  Elements $M\in SL(2,\RR)$ are called hyperbolic, elliptic and
parabolic, respectively, if the matrix trace satisfies $|\tr(M)|>2$,
$|\tr(M)|<2$ and $|\tr M|=2$. For elements in the image of the exponential map
this corresponds to $\bp^2<0$ ($\bp$ spacelike), $\bp^2>0$ ($\bp$ timelike) and
$\bp^2=0$ ($\bp$ lightlike).

The representation by $SO^+(2,1)$ matrices coincides with the adjoint action of
$SL(2,\RR)$ on its Lie algebra which is given by
  \begin{align}\label{addef}
    \tensor{\Ad(g)}{^c_a}J_c \defeq g\cdot J_a\cdot g^\inv\quad \forall g\in SL(2,\RR), %a\in\{0,1,2\},
    \qquad
    \tensor{\ad(J_a)}{^c_b}J_c\defeq \ad(J_a)(J_b) \defeq [J_a,J_b]% && \forall a,b\in\{0,1,2\}
    .
  \end{align}

Formula \eqref{lorbrac} for the Lie bracket then implies
$
  \ad(J_a)_{bc}=-\ee_{abc}$,
and the exponential map $\exp:\mathfrak{so}(2,1)\rightarrow SO^+(2,1)$ takes the
form
\begin{equation}\label{Ad-exp}
  \begin{split}
    \exp(\ad(p^cJ_c))_{bc} &= \Ad\bigl(\exp(p^aJ_a)\bigr)_{bc} \\
      &= \begin{cases}
        \hp_b\hp_c+\cos\mu\,(\eta_{bc}-\hp_b\hp_c)-\sin\mu\,\ee_{bcd}\,\hp^d & \text{if $\bp^2>0$},\\
        -\hp_b\hp_c+\cosh\mu\,(\eta_{bc}+\hp_b\hp_c)-\sinh\mu\,\ee_{bcd}\,\hp^d & \text{if $\bp^2<0$},\\
        \eta_{bc}+\frac12p_bp_c-\ee_{bcd}\,p^d & \text{if $\bp^2=0$}.
      \end{cases}
  \end{split}
\end{equation}
The  Poincar\'e group in three dimensions is the semidirect product
$P_3=SO^+(2,1)\ltimes\mathbb R^3$. In the following, we will also work with
its double cover $\tilde P_3=SL(2,\RR)\ltimes\mathbb R^3$, where $SL(2,\RR)$ acts on
$\RR^3$ via the adjoint representation \eqref{addef}. We parametrise their elements  as
\begin{equation}\label{pparam}
  (u,\ba)=\bigl(u, -\Ad(u)\bj\bigr)\quad\text{with}\quad
    \ba,\bj\in\RR^3,u\in SO^+(2,1) \text{ or } u\in SL(2,\RR).
\end{equation}
Their group multiplication laws then take the form
\begin{equation}\label{groupmult}
  (u_1,\ba_1)\cdot(u_2,\ba_2)
    =\bigl(u_1u_2,\ba_1+\Ad(u_1)\ba_2\bigr)
    =\bigl(u_1u_2,-\Ad(u_1u_2)[\bj_2+\Ad(u_2^\inv)\bj_1]\bigr).
\end{equation}

A basis of its Lie algebra $\mathfrak{iso}(2,1)$ is given by the basis $\{J_a\}_{a=0,1,2}$ of $\mathfrak{so}(2,1)$ together with a basis  $\{P_a\}_{a=0,1,2}$ of the abelian Lie algebra $\RR^3$. In
this basis, the Lie bracket takes the form
\begin{equation}\label{poincare}
  [J_a,J_b]=\tensor{\ee}{_a_b^c} J_c,\qquad
  [J_a,P_b]=\tensor{\ee}{_a_b^c} P_c,\qquad
  [P_a,P_b]=0.
\end{equation}
The Lie algebra $\mathfrak{iso}(2,1)$ has two quadratic Casimir elements which
are given by
\begin{equation}\label{casimirs}
  M=P_a\cdot P^a,\qquad S=P_a\cdot J^a+J_a\cdot P^a.
\end{equation}
In the parametrisation \eqref{pparam}, the exponential map $\exp_{P_3}:
\mathfrak{iso}(2,1)\to P_3$ is given by
\begin{equation}\label{expp3}
  \exp_{P_3}(p^aJ_a+k^aP_a)=\bigl(\exp(p^cJ_c), \bj\bigr)\quad
  \text{with } \bj=T(\bp)\bk,
\end{equation}
where $T(\bp): \RR^3 \to \RR^3$ is the invertible linear map
\begin{equation}\label{tdef}
  \begin{split}
  T(\bp)_{ab}^{\pm 1}
    %&=\sgn\bigl(\bp^2\bigr) \hp_a\hp_b+ s(\bp/2)^{\pm 1}\tensor{\Ad\bigl(\exp(\pm p^aJ_a/2)\bigr)}{_a^c}\bigl[\eta_{cb}-\sgn\bigl(\bp^2\bigr) \hp_c\hp_b\bigr]\\
    &=\begin{cases}
      \hp_a\hp_b+\bigl(\tfrac 2 \mu \sin\tfrac \mu 2\bigr)^{\pm 1} [\cos\tfrac \mu 2 (\eta_{ab}-\hp_a\hp_b)\mp\sin\tfrac \mu 2 \ee_{abc}\hp^c] & \text{if $\bp^2>0$},\\
      -\hp_a\hp_b+\bigl(\tfrac 2 \mu \sinh\tfrac \mu 2\bigr)^{\pm 1} [\cosh\tfrac\mu 2 (\eta_{ab}+\hp_a\hp_b)\mp\sinh\tfrac\mu 2 \ee_{abc}\hp^c] & \text{if $\bp^2<0$},\\
      \eta_{ab}\mp \tfrac 1 2 \ee_{abc}p^c & \text{if $\bp^2=0$}.
    \end{cases}
  \end{split}
\end{equation}

%%%%%%%%%%%%%%%%%%%%%%%%%%%%%%%%%%%

\subsection{Spacetime geometry in (2+1) dimensions}
\label{sec:3dgrav}

In the following we consider (2+1)-dimensional gravity with vanishing
cosmological constant on manifolds of topology $M\approx I\times S_{g,n}$, where
$I\subset\RR$ is an interval and $S_{g,n}$ an orientable surface of genus $g\geq
0$ with $n$ punctures. The punctures represent massive point particles with spin.

As the Ricci curvature tensor of a Lorentzian or Riemannian  three-dimensional manifold determines its Riemann
curvature tensor, solutions of the vacuum Einstein equations are of constant
curvature given by the cosmological constant. For the case
of vanishing cosmological constant this implies that any solution
of the vacuum Einstein equations is flat and locally isometric to
three-dimensional Minkowski space $\mathbb M^3$. The theory has no local
gravitational degrees of freedom but a finite number of global degrees of
freedom which arise from the topology and matter content of the spacetime.

\paragraph{Vacuum spacetimes}
Spacetimes that are solutions of the Einstein equations without matter are referred to as vacuum spacetimes, and their
mathematical properties are rather well-understood. The simplest spacetimes of this
type are maximal globally hyperbolic spacetimes whose Cauchy surface is
an orientable, compact surface $S_g$ of genus $g\geq 2$. These spacetimes have been classified
completely in \cite{Mess:2007aa}, for an overview over their geometrical properties see also \cite{Andersson:2007aa,Benedetti:2009aa}.
 They are of topology
$M\approx \RR\times S_g$ and are most easily described in terms of their universal cover.

In this description, the spacetimes are obtained as quotients of open, convex,
future complete regions in Minkowski space by a free and properly discontinuous
action of the fundamental group $\pi_1(S_g)$. This action of the fundamental
group is given by a group homomorphism $h:\pi_1(S_g)\rightarrow P_3$ which is
such that the Lorentzian components of its image form a cocompact Fuchsian group
$\Gamma\subset PSL(2,\RR)$ of genus $g$. In particular, this implies that the Lorentzian components of all holonomies  are hyperbolic.
 The action of the fundamental group induces a foliation of the
universal cover by surfaces of constant geodesic distance from the initial
singularity of its spacetime, which are preserved by the action of
$\pi_1(S_g)$. The spacetime is then obtained by identifying on each surface the
points that are related by this action of the fundamental group. It is shown in
\cite{Mess:2007aa} that the group homomorphism $h:\pi_1(S_g)\rightarrow P_3$ characterises
the spacetime completely and that two spacetimes constructed in this way are
isometric if and only if the associated group homomorphisms are related by
conjugation with $P_3$. This implies that the physical (gauge-invariant) phase
space of the theory is given by
\begin{align}\label{vacphsp}
  \gaugeinvpspace=\text{Hom}_0(\pi_1(S_g), P_3)/P_3,
\end{align}
where the index $0$ indicates the restriction to group homomorphisms whose
Lorentzian component $h:\pi_1(S_g)\rightarrow PSL(2,\RR)$ defines a cocompact Fuchsian
group of genus $g$.

\paragraph{Point particles}
A second class of well-known (2+1)-spacetimes are those with
particles. They were first investigated in the Euclidean context by
Staruszkiewicz \cite{Staruszkiewicz:1963aa} and for Lorentzian signature by Deser, Jackiw and 't Hooft
\cite{Deser:1984aa}, see also the later work \cite{Deser:1988aa,Sousa-Gerbert:1990aa,tHooft:1993aa,tHooft:1993ab,tHooft:1996aa}.

\begin{figure}
  \centering
  \includegraphics{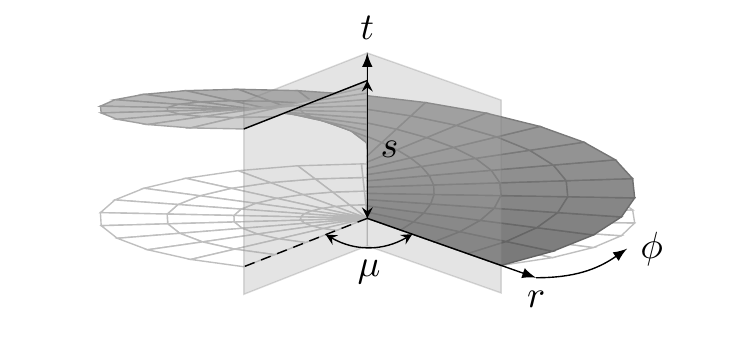}
  \caption{Spacetime for a single particle.  The metric for the spacetime with a point particle is obtained by cutting out a wedge in Minkowski space and identifying its boundary via a rotation by an angle $\mu$ and a translation $s$. The figure shows the wedge and a surface $(\tau=\text{const},\rho,\phi)$.}
  \label{fig:wedge}
\end{figure}

The simplest model is a spacetime containing a single point particle in $\RR^3$ that is at rest at
the origin.  The corresponding metric was constructed in \cite{Deser:1984aa}.  In cylindrical coordinates
$(\tau,\rho,\phi)$ it  takes the form
\begin{align}
\label{conem}
&ds^2=(d\tau+\tfrac s {2\pi}\,d\phi)^2-\frac{1}{(1-\tfrac{\mu}{2\pi})^2}d\rho^2-\rho^2
d\phi^2,
\end{align}
where $s\in\RR$ and $\mu\in[0,2\pi)$. The parameter $\mu$ can be
interpreted as the mass of the particle in units of the gravitational constant, while the parameter $s$ describes an
internal angular momentum or spin in units of $\hbar$.  It is shown in \cite{Deser:1984aa} that in terms of a new set of coordinates $( t,
r, \varphi)$ that are related to the cylindrical coordinates $(\tau,\rho,\phi)$ via
\begin{align}
t(\tau, \rho, \phi)=\tau+\tfrac s {2\pi}\, \phi,\qquad
 r(\tau, \rho, \phi)=\frac{\rho}{1-\tfrac\mu {2\pi}},\qquad
\varphi(\tau, \rho, \phi)=(1-\tfrac\mu {2\pi})\phi,
\end{align}
the metric \eqref{conem} takes the form of the Minkowski metric
\begin{equation}
  ds^2=d t^2-d r^2- r^2d \varphi^2.
\end{equation}
However, the range of the coordinate $\varphi$ is no longer $[0,2\pi)$ but $[0, 2\pi-
\mu)$. This implies that the resulting spacetime is locally, but not globally
isometric to Minkowski space. Instead of Minkowski space, the metric
\eqref{conem} describes a conical spacetime that is obtained by cutting out a
wedge of Minkowski space as shown in Figure \ref{fig:wedge} and
identifying its boundary according to the prescription $(t,r,0) \sim (t+ s, r, \mu)$.
This identification of the boundary is given by a rotation around the
$t$-axis by an angle $\mu$ combined with a translation in the direction of the $t$-axis by
a distance $s$.  The spacetime therefore has the geometry of a cone whose
deficit angle $\Delta\varphi=\mu$ and time shift $\Delta t=s$ are given by the
mass and spin of the particle.  It should be noted that in the case of non-vanishing spin the spacetime exhibits closed timelike curves. However, these curves are no longer present if a small region around each particle is excised from the spacetime \cite{Deser:1984aa,Carlip:2003aa}.

The construction generalises straightforwardly to a  point particle moving in
Minkowski space, whose worldline is a future directed, timelike geodesic.
This geodesic takes the form
\begin{align}\label{minkgeod}
g(T)=T\,\hat \bp+\bx, \qquad \bx\cdot \bp=0,
\end{align}
where $T$ is the eigentime associated with the particle, $\hat\bp$ is a timelike
unit vector that describes its velocity and $\bx$ is the offset of its
worldline from the origin.
In this case, the spacetime is obtained by cutting out a wedge  formed by two half-planes that
intersect in the particle's worldline.
 The identification of the boundary is  given by
the unique Poincar\'e transformation $M\in P_3$ that
 preserves the geodesic $g$, whose Lorentzian component is a rotation by an angle $\mu$ around $\hat\bp$
and whose translation
component in the direction of $\hat\bp$ is a translation by a distance $s$.  In
terms of the parametrisation \eqref{pparam}, it takes the form
\begin{align}\label{mpartpar}
&M=(u,-\Ad(u)\bj)=(v,\bx)\cdot (\exp(-\mu J_0), -s\, e_0)\cdot (v,\bx)^\inv,
\end{align}
with $u, v \in SO^+(2, 1)$ and $\bj, \bx \in \RR^3$.  The relation between the
variables $u, \bj$ and $v, \bx$ is given by
\begin{align}\label{ujexpress}
&u=v\exp(-\mu J_0) v^\inv=\exp(-\mu\, \hat p^c J_c), \qquad\bj=s\hat\bp+\idadi{}\bx.
\end{align}
The spacetimes associated with a particle of given mass $\mu$ and spin $s$ can
therefore be identified with a $P_3$-conjugacy class
\begin{equation}\label{conj0}
C_{\mu,s}=\{ g\cdot (\exp(-\mu J_0), s\be_0)\cdot g^\inv \mid g\in P_3\},
\end{equation}
or, equivalently, with an element $(\hat\bp, \bx)\in T^* \mathbb H^2$ of the
cotangent bundle of two-dimensional hyperbolic space.

The three-vector $\bp=\mu\,\hat\bp=\mu\Ad(v)e_0$ is the momentum three-vector
associated with the particle. It describes the particle's momentum and energy.
The three-vector $\bx$ describes the position of
the origin in the momentum rest frame of the particle.
The three-vector $\bj$
can be interpreted as a generalised angular momentum three-vector. Its component
in the direction of $\bp$ describes the internal angular momentum or spin $s$ of the
particle.  Its components orthogonal to $\bp$ describe an orbital angular
momentum arising from its motion in Minkowski space.

It is instructive to compare expression \eqref{ujexpress} for the angular
momentum $\bj$ in terms of the momentum and position variables $\bp$, $\bx$ with
the standard expression for a particle moving in Minkowski space. The latter is
given by
\begin{equation}\label{pjkdef}
  \bk=s\hat \bp+\bx\wedge\bp.
\end{equation}
A short calculation shows that the relation between the angular momenta $\bj$,
$\bk$ is given by the map $T(\bp): \mathbb R^3\rightarrow \mathbb R^3$ defined
in \eqref{tdef}.  We have $\bj=T(\bp)\bk$, and the expression \eqref{ujexpress}
for $\bj$ reduces to \eqref{pjkdef} in the limit $\mu\rightarrow 0$. A detailed
discussion of this relation and its physical interpretation is given in \cite{Meusburger:2003aa}.

For the following, it will be useful to consider generalised cones associated
with spacelike or timelike geodesics. The construction is analogous to the
timelike case. Again, the associated spacetimes are obtained by removing a wedge
from Minkowski space and identifying the boundary of the resulting region. The
geodesic that describes the intersection of the associated planes takes the form
\eqref{minkgeod}, but the vector $\hat\bp$ is replaced by a spacelike unit
vector or by a lightlike vector. The Poincar\'e transformation that describes the
identification of points of the boundary takes the form
\begin{align}\label{paramgeom}
  M=(\exp(-p^cJ_c), -\Ad(\exp(-p^cJ_c))\bj)
\end{align}
with $\bp,\bj \in\RR^3$. In the spacelike case, the vector $\bp$ describes a
Lorentz boost with rapidity $\mu=\sqrt{|\bp^2|}$, in the lightlike case it
describes a parabolic element of $PSL(2,\RR)$.

\paragraph{Multi-particle spacetimes}
Spacetimes with multiple point particles were first introduced in \cite{Staruszkiewicz:1963aa,Deser:1984aa,Deser:1988aa}. Their
physical properties are well-understood and have been studied extensively in the physics
literature \cite{Staruszkiewicz:1963aa,Deser:1984aa,Deser:1988aa,Sousa-Gerbert:1990aa,tHooft:1993aa,tHooft:1993ab,tHooft:1996aa}, for an overview see \cite{Carlip:2003aa}. However, their mathematical structure is considerably more
involved than that of vacuum spacetimes, and a systematic investigation of their
mathematical features has been initiated only recently \cite{Krasnov:2007ys,Bonsante:2010rt,Barbot:2010di}.

Spacetimes with particles define
manifolds with conical singularities or, in the case where the masses of all
particles are rational multiples of $2\pi$, orbifolds.
With the exception of the orbifold case, they cannot be obtained as quotients of regions in three-dimensional
Minkowski space. This is due to the fact that the elliptic elements of
$PSL(2,\RR)$ associated with the particles do not give rise to a free and
properly discontinuous action of the fundamental group on hyperbolic space
$\mathbb H^2$ and the developing map is no longer  an embedding \cite{Bonsante:2010rt,Barbot:2010di}.

Although a classification and explicit description of such spacetimes is still missing, examples  can be constructed by gluing the boundary of certain
regions $D\subset \mathbb M^3$ in Minkowski space.
The
boundary of the region $D\subset\mathbb M^3$ can be divided into components
which are identified pairwise by certain Poincar\'e transformations.
This identification is given by a group homomorphism $h:
\pi_1(S_{g,n})\rightarrow P_3$ from the fundamental group of the spatial surface
$\surf$ into the Poincar\'e group and can be specified through the images of a set of generators.

\begin{figure}
  \centering
  \includegraphics{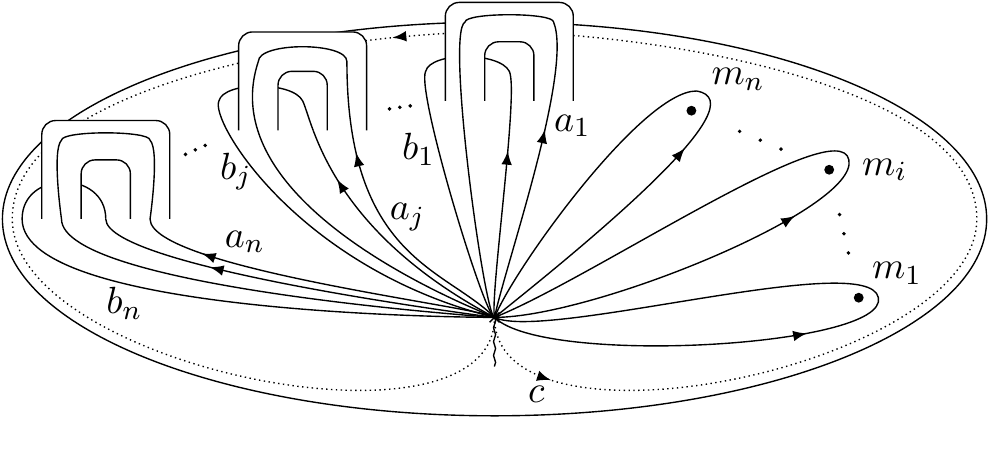}
  \caption{Generators of the fundamental group for an $n$-punctured genus $g$ surface $\surf$.  The
    generators of the fundamental group $\pi_1(\surf)$ are the homotopy equivalence  classes of the curves $m_1, \dots,
    m_n, a_1, b_1, \dots, a_n, b_n$.  The defining relation of $\pi_1(\surf)$
    states that the curve $c$ is  contractible. The short wavy line indicates the cilium that defines a linear ordering of the incident edges at the basepoint as explained in Section~\ref{sec:fock-rosly}.}
  \label{fig:fundamental-group}
\end{figure}

\begin{figure}
  \centering
  \includegraphics{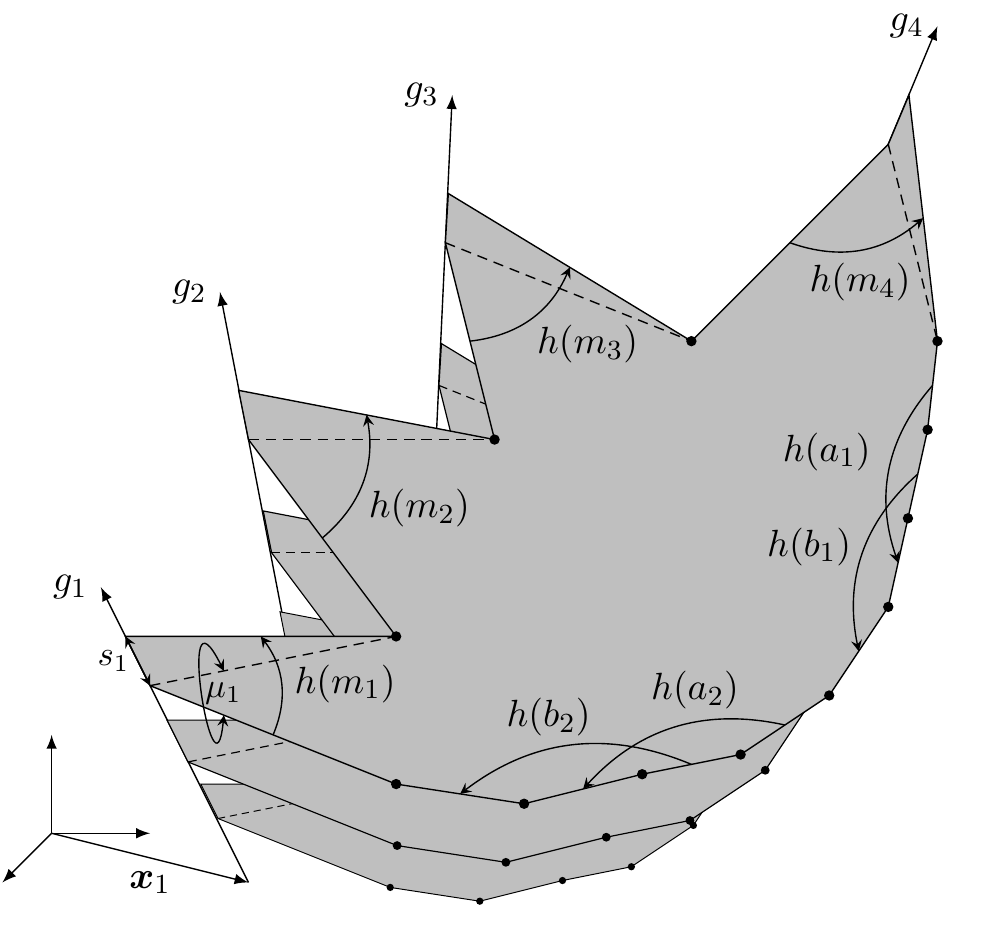}
  \caption{Identification of the boundary for a region  $D \subset \mathbb M^3$ for a surface $S_{2,4}$ with four punctures and two handles.}
  \label{fig:domain}
\end{figure}

The fundamental group $\pi_1(S_{g,n})$ of a
genus $g$ surface with $n$ punctures is generated by the homotopy equivalence
classes of loops $m_i$, $i=1,\dots,n$, around each puncture and by the $a$- and
$b$-cycles $a_j,b_j$, $j=1,\dots,g$, associated to each handle as depicted in
Figure \ref{fig:fundamental-group}. It has a single defining relation, which states that the curve $c$ depicted in
Figure \ref{fig:fundamental-group} is contractible:
\begin{equation}\label{pi1rel}
  [b_g,a_g^\inv]\circ \ldots\circ[b_1,a_1^\inv]\circ m_n\circ \ldots\circ m_1 = 1,\qquad
  [b_g,a_g^\inv] \defeq b_g\circ a_g^\inv\circ b_g^\inv\circ a_g.
\end{equation}
The identification of the boundary of  $D\subset\mathbb M^3$ is thus given by a set of Poincar\'e elements $M_i$, $A_j$, $B_j\in P_3$ that satisfy a relation analogous to \eqref{pi1rel} and which are such that the elements $M_i$ are restricted to fixed $P_3$ conjugacy classes \eqref{conj0}. They identify the
 boundary components
of the region pairwise as shown in Figure \ref{fig:domain}.  The Poincar\'e elements
$M_i\in \mathcal  C_{\mu_i,s_i}$ identify
two adjacent
segments that intersect in a timelike geodesic stabilised by $M_i$. This
geodesic corresponds to the worldline of the associated particle. The Lorentzian components of the elements $M_i$ are therefore elliptic.
The elements $A_j,B_j$ identify the remaining sides  as shown in  Figure \ref{fig:domain}.
They correspond to the handles on the surface $\surf$. Their Lorentzian components are therefore hyperbolic.

As the Poincar\'e group $P_3$ is the  group of orientation and time orientation preserving isometries of Minkowski space, the
resulting three-manifold with singularities inherits a flat Lorentz metric
induced by the Minkowski metric. As it is an isometry, applying a global Poincar\'e
transformation to the associated region $D\subset \mathbb M^3$ does not affect
the geometry of the resulting spacetime. This corresponds to a transformation of
the associated group homomorphism $h: \pi_1(S_{g,n})\rightarrow P_3$ by
conjugation.  In terms of the variables $\bp$ and $\bj$ defined in \eqref{paramgeom}, conjugation of a holonomy by a Poincar\'e element $(w,\by)$ corresponds to a transformation
\begin{align}\label{pjconj}
\bp\rightarrow \Ad(w)\bp, \qquad \bj\rightarrow \Ad(w)\bj+[\mathds{1}-\Ad(w\exp(p^cJ_c) w^\inv)]\by.
\end{align}
The transformation of the variables $\bp,\bj$ under a Lorentz transformation $w\in PSL(2,\RR)$ is therefore given by
\begin{align}\label{ltrnsf}
\bp\rightarrow \Ad(w)\bp, \qquad \bj\rightarrow \Ad(w)\bj,
\end{align}
and their transformation under a translation $\by\in\RR^3$ takes the form
\begin{align}\label{ttrnsf}
\bp\rightarrow \bp,\qquad \bj\rightarrow \bj+[\mathds{1}-\Ad(\exp(p^aJ_a))]\by.
\end{align}

As in the vacuum case, spacetimes whose associated group
homomorphisms are related by global conjugation therefore should be identified.
However, in contrast to the vacuum case, little is known about the properties of the associated
regions $D\subset \mathbb M^3$.  For instance, for the
case of point particles on a sphere, there exist non-isometric spacetimes that
correspond to the same group homomorphism $h:\pi_1(S_{g,n})\rightarrow P_3$ \cite{Matschull:1999dq,Benedetti:2009aa}. This phenomenon is known under the name of ``holonomy failure'' in mathematics.

In the following we focus on the formulation of (2+1)-gravity as a Chern-Simons
gauge theory \cite{Achucarro:1986aa,Witten:1988aa}. This is a closely related, but non-equivalent formulation of
the theory whose phase space structure is well-understood. Our results do
therefore not make use of the geometrical description of these spacetimes in
terms of the universal cover and do not rely on geometrical classification
results. However,  the geometrical description will provide us with physical
intuition about the spacetimes and act as a guideline for the physical
interpretation of our results.

%%%%%%%%%%%%%%%%%%%%%%%%%%%%%%%%%%%

\subsection{The Chern-Simons formulation of (2+1)-gravity}
\label{sec:cs}

The Chern-Simons formulation of (2+1)-gravity \cite{Achucarro:1986aa,Witten:1988aa} is derived from Cartan's
formulation of the theory in terms of a triad $e=e_\mu dx^\mu$ and spin
connection $\omega=\omega_\mu dx^\mu$ on a smooth three-manifold $M$. It is obtained by combining the
triad and spin connection into a Chern-Simons gauge field
\begin{equation}
  A=\omega^a J_a+e^aP_a,
\end{equation}
where $J_a$, $P_a$ denote the generators \eqref{poincare} of $\mathfrak{iso}(2,1)$. The gauge field is a connection 1-form of a $P_3$ principal bundle on $M$.  It determines the metric
via
\begin{equation}\label{metrictriad}
  g_{\mu\nu}=\eta_{ab}e^a_\mu e^b_\nu.
\end{equation}
It is shown in \cite{Witten:1988aa} that the Einstein-Hilbert action associated with Cartan's
formulation of the theory can be expressed as a Chern-Simons action
\begin{equation}\label{csact}
  S[A]=\frac k{4\pi}\int_M \tr(A\wedge \diffd A+\tfrac 2 3 A\wedge A\wedge A),
\end{equation}
where $k=1/4G$ and where the trace $\tr(ST) = \langle S, T \rangle$ is given in
terms of the $\Ad$-invariant symmetric bilinear form on $\mathfrak{iso}(2,1)$
defined by
\begin{equation}\label{form}
  \langle J_a,J_b\rangle=\langle P_a,P_b\rangle=0, \qquad
  \langle J_a,P_b\rangle=\eta_{ab}.
\end{equation}
It is shown in \cite{Witten:1988aa} that the transformations of triad and spin connection under infinitesimal diffeomorphisms are on-shell equivalent to infinitesimal Chern-Simons
gauge transformations
\begin{equation}
  A\mapsto \gamma A \gamma^\inv + \gamma \, \diffd\gamma^\inv, \qquad
  \gamma: M\rightarrow P_3.
\end{equation}
Note, however, that this equivalence between gauge transformations and
diffeomorphisms does not hold for large, \ie not infinitesimally generated,
gauge transformations and diffeomorphisms. Moreover,  to define a non-degenerate metric via
\eqref{metrictriad}, the triad is required to
be non-degenerate in Cartan's formulation, whereas no such condition is imposed in the Chern-Simons
formulation. It is shown in \cite{Matschull:1999dq} (for a discussion of the
two-dimensional case see also \cite{Schaller:1994aa}) that this leads to discrepancies between
the phase spaces of the two theories.

Point particles are included into the formalism via  minimal coupling  to the Chern-Simons gauge field \cite{Sousa-Gerbert:1990aa}. This requires the choice of a
coadjoint orbit of the gauge group for each of the punctures which are determined by elliptic elements
\begin{equation}\label{specels}
  D_{\mu_i,s_i}=\mu_i J_0+s_i P_0 \in \mathfrak{iso}(2,1).
\end{equation}
As suggested by the notation, the parameters $\mu_i,s_i$ define the masses and spins of the associated particles. The Chern-Simons
action with point particles then depends on these parameters as well as
dynamical variables $h_i\in P_3$, $i=1,\dots,n$, associated to each puncture.
%In terms of these variables.
%, it takes the form
%\begin{multline}\label{actpart}
%S[A,\{h_i\}]=\frac k {4\pi}\int_M \biggl\{\langle A\wedge \diffd A+\tfrac 2 3 A\wedge A\wedge A\rangle\\
 %-\sum_{i=1}^n \langle D_{\mu_is_i}, h_i^\inv \diffd h_i\rangle-\sum_{i=1}^n \int_M \langle  h_i D_{\mu_i,s_i} h_i^\inv \delta(x-x_{(i)}), A\rangle \biggr\},
%\end{multline}
The equations of motion derived from this action are a condition on the
curvature and on the value of the gauge field at the punctures:
\begin{equation}\label{curvcond}
\begin{gathered}
F=\diffd A+A\wedge A=\frac k {4\pi}\sum_{i=1}^n h_iD_{\mu_i,s_i} h_i^\inv\,\delta(x-x_{(i)}),\\
A(x_{(i)})=h_i D_{\mu_i s_i} h_i^\inv +h_i \diffd h_i^\inv,
\end{gathered}
\end{equation}
where the coordinate $x_{(i)}$ parametrises the worldline of the puncture on
$M$.  This condition
 forces the gauge field $A$ to be flat everywhere outside the worldlines of
the particles and to take a fixed form in the vicinity of these worldlines. As the curvature in
\eqref{curvcond} combines the curvature and the torsion from Cartan's
formulation
\begin{equation}
F=F_\omega^a J_a+T^aP_a\quad\text{with}\quad
F_\omega^a=\diffd\omega^a+\tfrac 1 2\ee^{abc}\omega_b\omega_c , \quad T^a=\diffd e^a+\ee^{abc}\omega_be_c,
\end{equation}
the masses $\mu_i$ appear as  sources of curvature and the spins
$s_i$ as  sources of torsion.

On manifolds of topology $I\times S_{g,n}$, where $I\subset\RR$ is an interval and $S_{g,n}$
a compact orientable surface with punctures, one can give a Hamiltonian
formulation of the theory which is obtained by splitting the gauge field as
$A=A_0(x^0,x_S) \diffd x^0+A_s(x^0,x_S)$, where $x_S=(x_1,x_2)$ are the
coordinates that parametrise the surface $S_{g,n}$ and $x^0$ parametrises $I$. The
associated equations of motion are a flatness condition similar to
\eqref{curvcond} on the spatial gauge field $A_s$ and a set of evolution
equations, which state that the evolution of the system with respect to the parameter $x^0$
is pure gauge.

It follows from these considerations \cite{Witten:1988aa,Sousa-Gerbert:1990aa} that
 the gauge-invariant phase space of the theory is the moduli
space of flat $P_3$-connections modulo gauge transformations on the (punctured)
surface $S_{g,n}$. It is is given as  the set of conjugation equivalence classes
\begin{equation}\label{pnctphsp}
  \gaugeinvpspace=\text{Hom}_{(\mu_1,s_1),\dots,(\mu_n,s_n)}(\pi_1(S_{g,n}), P_3)/P_3
\end{equation}
of group homomorphisms $h:\pi_1(S_{g,n})\rightarrow P_3$ which map the images of the
loops around each puncture to fixed conjugacy classes
\begin{equation}\label{conj}
  \mathcal C_{\mu_i,s_i}=\{h\cdot\exp(\mu_i J_0+s_i P_0)\cdot h^\inv \mid h\in P_3\}.
\end{equation}
Using the set of generators of the fundamental group $\pi_1(S_{g,n})$ introduced in Section
\ref{sec:3dgrav}, we can characterise a group homomorphism $h: \pi_1(\surf)\rightarrow P_3$
through the images of the generators $m_i$, $a_j$, $b_j$.
 The set of group homomorphisms $h:\pi_1(S_{g,n})\rightarrow
P_3$ which is such that the images of the loops $m_i$ lie in the conjugacy
classes \eqref{conj} can then be identified with the submanifold
\begin{multline}\label{holpi1desc}
  \{(M_1,\dots,M_n,A_1,B_1,\dots,A_g,B_g)\in P_3^{n+2g} \mid \\
    M_i\in\mathcal C_{\mu_i, s_i}, \, [B_g,A_g^\inv] \cdot [B_1,A_1^\inv] \cdot M_n\cdots M_1=1\}
  \subset P_3^{n+2g},
\end{multline}
and the gauge-invariant phase space $\gaugeinvpspace$ in \eqref{pnctphsp} is
obtained from this manifold by identifying points which are related by global
conjugation with $P_3$:
\begin{equation}\label{phasehol}
  \gaugeinvpspace = \{(M_1,\dots,B_g)\in P_3^{n+2g} \mid
    M_i\in\mathcal C_{\mu_i, s_i}, [B_g,A_g^\inv] \cdot [B_1,A_1^\inv] \cdot M_n\cdots M_1=1\}/P_3.
\end{equation}

In the Chern-Simons formalism, the images of elements of the fundamental group under such group homomorphisms
are obtained as the path-ordered exponential of the gauge field along closed loops on $S_{g,n}$. As
the gauge field satisfies the flatness condition \eqref{curvcond}, the value of
such path-ordered exponentials depends only on the homotopy equivalence class of
the loop, and it transforms by conjugation under a change of the basepoint. In
the following we will refer to these path-ordered exponentials as holonomies.

Note that the holonomies are directly related to the corresponding group
homomorphisms in the geometrical picture (\cf Section \ref{sec:3dgrav}). The
restriction of the holonomies associated with loops around the particles to
conjugacy classes \eqref{conj} mirrors the corresponding condition \eqref{conj0} in the
geometrical description. For the case without punctures, expression
\eqref{pnctphsp} for the phase space in the Chern-Simons formalism is directly
related to the corresponding expression \eqref{vacphsp} in the metric
formulation. Note, however, that the restriction to group homomorphisms that
define cocompact Fuchsian groups of genus $g$ is absent in the Chern-Simons
formulation. This reflects the presence of configurations corresponding to
degenerate metrics.

%%%%%%%%%%%%%%%%%%%%%%%%%%%%%%%%%%%

\subsection{Phase space and Poisson structure}
\label{sec:fock-rosly}

The moduli space of flat connections  \eqref{pnctphsp}  is equipped with a canonical symplectic structure obtained via
symplectic reduction from the canonical symplectic structure associated with the
Chern-Simons action.  The description \eqref{pnctphsp} of the phase space in
terms of group homomorphisms $h: \pi_1(S_{g,n})\rightarrow P_3$ has the
advantage that it gives rise to an efficient and explicit parametrisation of
this canonical symplectic structure on an ambient space.  This description is
due to Alekseev and Malkin \cite{Alekseev:1995aa} and Fock and Rosly \cite{Fock:1998aa} and has served as a
central ingredient of the combinatorial quantisation formalism for Chern-Simons
gauge theory \cite{Alekseev:1995ab,Alekseev:1996aa,Buffenoir:1995aa} and in its application to (2+1)-gravity \cite{Buffenoir:2002aa,Meusburger:2003hc,Noui:2006ku,Meusburger:2008aa,Meusburger:2010bc}.

%\subsubsection{Fock and Rosly's Poisson structure}

In the following, we will work with Fock and Rosly's description \cite{Fock:1998aa} of the moduli space.
It describes the canonical symplectic
structure on $\gaugeinvpspace$ in \eqref{pnctphsp} in terms of an auxiliary
Poisson structure on the group $P_3^{n+2g}$, where the different copies of $P_3$
stand for the holonomies along a set of generators of the fundamental group
$\pi_1(\surf)$ as in \eqref{holpi1desc}.

This Poisson structure is non-canonical, as  it
requires two additional ingredients in addition to the underlying gauge theory:
The first is a linear ordering of the edge ends incident at the basepoint of
  the fundamental group $\pi_1(\surf)$. As the orientation of the surface
  induces a cyclic ordering of these edges, a linear ordering is obtained by
  inserting a cilium that separates the edges of maximal and minimal order as
  shown in Figure~\ref{fig:fundamental-group}.

 The second ingredient is a classical $r$-matrix for the group $P_3$, \ie a solution of the classical Yang-Baxter equation, whose
  symmetric component is dual to the $\Ad$-invariant symmetric bilinear form
  \eqref{form} in the Chern-Simons action.
It is shown in \cite{Fock:1998aa} that, when reduced to the moduli space of flat
connections \eqref{phasehol}, this Poisson structure induces the canonical
symplectic structure on $\gaugeinvpspace$ for all choices of classical
$r$-matrices and of the ordering.

In the following, we use the presentation of the fundamental group given in Section
\ref{sec:3dgrav} and the set of representatives depicted in Figure \ref{fig:fundamental-group}.
Our choice of ordering of edge ends at the basepoint is the one depicted in Figure
\ref{fig:fundamental-group}. It defines the following  partial ordering of the holonomies
$M_1,\dots,M_n,A_1,B_1,\dots,A_g,B_g$:
\begin{equation}\label{ordering}
  X < Y \Leftrightarrow
  \begin{cases}
    X=M_i, Y=M_j \text{ with } i<j \text{, or } \\
    X\in\{M_1,\dots,M_n\}, Y\in\{A_1,B_1,\dots,A_g,B_g\} \text{, or } \\
    X\in\{A_i,B_i\}, Y\in\{A_j,B_j\} \text{ with } i<j.
  \end{cases}
\end{equation}
Our choice of the classical $r$-matrix is the one that corresponds to the
Drinfel'd double $DSO(2,1)$ and is given in terms of the generators $P_a,J_a$ in
\eqref{poincare} as $r=P_a\oo J^a$.  Fock and Rosly's Poisson structure \cite{Fock:1998aa} for these conventions
 has been derived in \cite{Meusburger:2003aa}, see also \cite{Meusburger:2008aa}. The resulting
description  is summarised in the following theorem.

\begin{theorem}[\cite{Meusburger:2003aa}]\label{thm:frtheorem}
In terms of the coordinates
\begin{equation}\label{coords}
  p^b: (\exp(-q^cJ_c), \ba)\mapsto q^b,\qquad
  j^b: (\exp(-q^cJ_c), \ba)\mapsto -\tensor{\Ad(\exp(+q^cJ_c))}{^b_d}\,a^d,
\end{equation}
on the different copies of $P_3$, Fock and Rosly's Poisson
structure \cite{Fock:1998aa} on $P_3^{n+2g}$ is given by
\begin{subequations}\label{frbracket}
\begin{align}
  \label{singlebr}&\left.\begin{aligned}
    \{p_\mi^a,p_\mi^b\}&=0\\
    \{j_\mi^a,p_\mi^b\}&=-\tensor{\ee}{^{ab}_c} \ p^c_\mi\\
    \{j_\mi^a,j_\mi^b\}&=-\tensor{\ee}{^{ab}_c} \ j^c_\mi
  \end{aligned}\hspace{11.3em}\,\right\}\quad i=1,\ldots,n, \\[1em]
  \allowpagebreakhere
  \label{mixbr}&\hspace{0.75em}\left.\begin{aligned}
    \{j_{X}^a,p_{Y}^b\}&=-\tidadi{X}{^a_d} \ \tensor{\ee}{^{db}_c} \ p^c_{Y}\\
    \{j_{X}^a,j_{Y}^b\}&=-\tidadi{X}{^a_d} \ \tensor{\ee}{^{db}_c} \ j^c_{Y}\\
    \{p_{X}^a,j_{Y}^b\}&=0\\
    \{p_{X}^a,p_{Y}^b\}&=0
  \end{aligned}\hspace{4.5em}\right\}\quad X<Y, \\[1em]
  \allowpagebreakhere
  \label{handlebr}&\hspace{0.4em}\left.\begin{aligned}
    \{p_\ai^a,p_\ai^b\}&=0\\
    \{j_\ai^a,p_\ai^b\}&=-\tensor{\ee}{^{ab}_c} \ p^c_\ai\\
    \{j_\ai^a,j_\ai^b\}&=-\tensor{\ee}{^{ab}_c} \ j^c_\ai\\
    \{p_\bi^a,p_\bi^b\}&=0  \\
    \{j_\bi^a,p_\bi^b\}&=-\tensor{\ee}{^{ab}_c} \ p^c_\bi\\
    \{j_\bi^a,j_\bi^b\}&=-\tensor{\ee}{^{ab}_c} \ j^c_\bi\\
    \{j_\ai^a,j_\bi^b\}&=-\tensor{\ee}{^{ab}_c} \ j^c_\bi\\
    \{p_\ai,p_\bi\}&=0\\
    \{j_\ai^a,p_\bi^b\}
      %&=-\tensor{\left(1+\Ad({u_\ai^\inv})
       % \cdot\frac{\Ad(u_\bi)}{1-\Ad(u_\bi)}\right)}{^a_d}
        %\ \ee^{bcd} \ p_c^\bi\\
      &=-\tensor{\ee}{^{ab}_c} \ p^c_\bi+\tensor{\Ad(u^\inv_\ai)}{^a_c} \ {T^\inv(\bp_\bi)}^{cb}\\
    \{j_\bi^a,p_\ai^b\}
      %&=\tensor{\left(\frac{\Ad(u_\ai)}{1-\Ad(u_\ai)}\right)}{^a_d}
        %\ \ee^{bcd} \ p_c^\ai\\
      &=-T^\inv(\bp_\ai)^{ab}
  \end{aligned}\quad\right\}\quad i=1,\ldots,g,
\end{align}
\end{subequations}
where the linear map $T(\bp): \RR^3\rightarrow \RR^3$ and its inverse
$T(\bp)^\inv$ are given by \eqref{tdef} and the notation $X<Y$ refers to the partial ordering \eqref{ordering}.
\end{theorem}

It is shown in \cite{Meusburger:2003aa, Meusburger:2003hc} that Fock and Rosly's Poisson algebra can be identified
with the Poisson algebra spanned by functions and vector fields on $n+2g$ copies
of  $SO^+(2,1)\cong PSL(2,\RR)$.  For this, one labels the
components of $PSL(2,\RR)^{n+2g}$ with $u_{M_1},\dots, u_{M_n}$, $u_{A_1}, u_{B_1},\dots,u_{A_g}, u_{B_g}$ and
denotes for $X\in\{M_1,\dots,M_n,A_1,B_1,\dots,A_g,B_g\}$ by $R^a_X,L^a_X$ the
associated left- and right-invariant vector fields on $PSL(2,\RR)^{n+2g}$,
\begin{equation}\label{vecfields}
  \begin{aligned}
  R^a_X f(u_{M_1},\dots,u_{M_n}, u_{A_1},\dots,u_{B_g})
    &=\tdiff{}{t}\bigg|_{t=0} f(u_{M_1},\dots, u_X\cdot e^{tJ_a},\dots,u_{B_g}),\\
  L^a_X f(u_{M_1},\dots,u_{M_n}, u_{A_1},\dots,u_{B_g})
    &=\tdiff{}{t}\bigg|_{t=0} f(u_{M_1},\dots, e^{-tJ_a}\cdot u_X,\dots,u_{B_g}).
  \end{aligned}
\end{equation}
The variables $j^a_X$ can then be identified with the following vector fields on $PSL(2,\RR)^{n+2g}$:
\begin{align}\label{jvecfields}
  j^a_\mi&=-(R^a_\mi+L^a_\mi)
    -\tidad{\mi}{^a_b}
     \smashoperator{\sum_{X>\mi}} \bigl( L^b_X+R^b_X \bigr),\\
  j^b_\ai&=-(R^a_\ai+L^a_\ai)-(R^a_\bi+L^a_\bi)
    -\tensor{\Ad(u_\ai^\inv u_\bi)}{^a_b}R^b_{\bi}
    -\tidad{\ai}{^a_b}
      \smashoperator{\sum_{X>\ai}} \bigl(L^b_X+R^b_X\bigr), \nonumber\\
  j^b_\bi&=-(R^a_\bi+L^a_\bi)-L^a_\ai
    -\tidad{\bi}{^a_b}
      \smashoperator{\sum_{X>\bi}} \bigl(L^b_X+R^b_X\bigr),\nonumber
\end{align}
where the expressions   $X>M_i,X>A_i,X>B_i$ refer to the partial ordering \eqref{ordering}.
The variables $p^a_X$ correspond to functions on $SL(2,\RR)^{n+2g}$. The Poisson
brackets of variables $j^a_X$ and $p^b_Y$  describe the action of the
associated vector fields on functions and the Poisson brackets between variables
$j^a_X$, $j^b_Y$ are given by the Lie bracket of the associated vector fields.

A detailed analysis of Fock and Rosly's Poisson structure for the group $P_3$ is
given in \cite{Meusburger:2003aa,Meusburger:2003hc}. It is shown there that it restricts to a symplectic structure
on the submanifold $\mathcal C_{\mu_1,s_1}\times\dots\times\mathcal
C_{\mu_n,s_n}\times P_3^{2g}$.
%This symplectic structure is Poisson-isomorphic to the cotangent
%bundle $T^*(\mathcal C_{\mu_1}\times\ldots\times\mathcal C_{\mu_N}\times
%PSL(2,\RR)^{2g})$, where $\mathcal C_\mu$ denotes the conjugacy class $C_{\mu}=\{g\cdot
%e^{\mu J_0}\cdot g^\inv \mid g\in SO(2,1)\}$.
Moreover, the results of \cite{Meusburger:2005wk} demonstrate
that the mapping class group $\text{Map}(\surf\setminus D)$ of the associated
surface $\surf\setminus D$ with a disc removed acts on this Poisson manifold by
Poisson isomorphisms. As different orderings of the particles and handles are
related by  the action of elements of the mapping class group, this implies in particular that
Fock and Rosly's auxiliary bracket \cite{Fock:1998aa} depends only trivially on the choice of the
ordering.

%%%%%%%%%%%%%%%%%%%%%%%%%%%%%%%%%%%

\section{Gauge fixing in (2+1)-gravity}
\label{sec:gauge}

\subsection{Constrained systems and gauge fixing}
\label{sec:gfixing}

In the following, we will consider Fock and Rosly's Poisson algebra as a constrained mechanical system  and construct the associated Dirac bracket.  We start by summarising
the relevant concepts from the theory of constrained systems \cite{Dirac:1950aa,Dirac:1951aa,Dirac:2001bh} following
\cite{Henneaux:1994aa}. (For a pedagogical introduction, see also
\cite{Matschull:1996aa} and \cite{Figueroa-OFarrill:1989aa}.)

A constrained mechanical system  is given by a Poisson manifold
$(\phasespace,\{\,\,\})$ together with a set of constraint functions
$\{\phi_i\}_{i=1,\dots,k}\subset\cif(\phasespace)$. The manifold
$(\phasespace,\{\,\,\})$ is called the extended phase
space of the theory. The constraint surface $\Sigma$ is defined by the condition $\csurface \defeq
\{q \in \phasespace \mid \phi_i(q) = 0 \ \forall i=1,\dots,k\}$.

Constraints can be classified via the concept of weak equality. Two
functions $F, G \in \cif(\phasespace)$ are said to agree weakly, $F \approx G$,
if they are equal on the constraint surface: $F(p)=G(p)$ for all
$p\in\csurface$.  This property implies that they differ by a linear combination of the constraints with $\cif(\phasespace)$-valued coefficients:
\begin{equation}
  F \approx G \Longleftrightarrow F - G = c^i \phi_i
    \text{ for some $c^i \in \cif(\phasespace)$}.
\end{equation}

First-class constraints are constraints whose Poisson bracket with all other
constraints vanishes weakly:
\begin{equation}
  \phi_i \text{ first-class}
    :\Leftrightarrow \{\phi_i, \phi_j\} \approx 0 \quad\forall j\in\{1,\dots,k\}.
\end{equation}
Constraints which are not first-class are called second-class.  Via the Poisson
bracket, a first-class constraint $\phi_i$ generates a gauge transformation
\begin{equation}\label{general-gauge-transformations}
  \delta_i F \defeq \{F, \phi_i\} = \tdiffat{}{t}{t=0} F \circ \varphi^t_i\quad \forall F \in \cif(\phasespace),
\end{equation}
where $\varphi_i: \RR \times \phasespace \to \phasespace$ denotes the flow
generated by $\phi_i$.  Formally, Casimir functions, \ie functions whose
Poisson bracket with all functions in $\cif(\phasespace)$ vanishes, are
also first-class.  However, as the associated gauge transformations are trivial,
we will not consider them to be first-class constraints in the following.

Functions $O \in \cif(\csurface)$ on the constraint surface that are weakly
invariant under gauge transformations are called observables:
\begin{equation}\label{observable-condition}
  O \in \cif(\csurface) \text{ observable}
    :\Leftrightarrow \delta_i O \approx 0 \quad \forall i \in \{1, \dots, l\}.
\end{equation}
Observables can also be thought of as equivalence classes of gauge-invariant
smooth functions on the extended phase space $\phasespace$, with the equivalence
relation given by $F \sim G :\Leftrightarrow F \approx G$.  Since for any
representative $O'$ in the equivalence class of $O$ we have $\delta_i O' \approx
\delta_i O$, the condition \eqref{observable-condition} is still well-defined in
this picture.

Observables encode the physical degrees of freedom of the system and describe possible outcomes of measurements.
In contrast,  the description of the system in terms of the extended phase space and the constraint surface is redundant. Both contain unphysical (``gauge'') degrees of freedom. This redundancy of the description is encoded in the gauge transformations. Any two points of the constraint surface which are related by the flows as in \eqref{general-gauge-transformations} describe the same physical state. A physical state thus corresponds to a gauge equivalence class of points on the constraint surface, and the outcome of measurements is described by gauge equivalence classes of functions on the extended phase space.

The physical content of the theory can therefore be described unambiguously by
restricting attention to the gauge-invariant observables. However, in practice
this is often not feasible as the resulting Poisson algebra can be
complicated. (For the case of (2+1)-gravity, see \cite{Nelson:1989aa,Nelson:1991aa,Nelson:1992aa}). An alternative approach is to construct a set of representatives for
the gauge equivalence classes of functions on the extended phase space
$\phasespace$. This is done via a gauge fixing procedure. It amounts to imposing an
additional set of constraints  $\chi_j\approx 0$,
$\chi_j\in\cif(\phasespace)$,  called gauge fixing conditions,
such that the gauge freedom generated by
first-class constraints is eliminated.

For simplicity, we restrict our discussion of the gauge fixing procedure to
constrained systems with only first-class constraints. Moreover, we suppose that
the constraint functions are non-redundant, \ie that the constraint surface
$\csurface$ is a submanifold of the $n$-dimensional manifold $\phasespace$ of
dimension $m=n-k$ and that the gradients $\text{grad}(\phi_i)$ of the constraint
functions are linearly independent everywhere on $\csurface$.  In this setting,
a good set of gauge fixing conditions must have two properties:
\begin{enumerate}
\item It must be accessible without changing the values of observables, \ie it
  must be possible to map any point $q \in \csurface$ on the constraint surface
  to one that satisfies the gauge fixing conditions by using the flows
  associated to the gauge transformations \eqref{general-gauge-transformations}.
\item It needs to eliminate the gauge freedom completely, \ie no gauge
  transformation may preserve all the gauge fixing conditions.  In other words,
  the number of gauge fixing conditions must agree with the number of
  constraints, and the matrix $C=(\{\chi_j, \phi_i\})_{i,j=1,\dots,k}$ must be
  invertible everywhere on the constraint surface.
\end{enumerate}
These conditions  imply that the gauge fixing conditions
$\{\chi_j\}_{j=1,\dots,k}$ together with the original constraints
$\{\phi_i\}_{i=1,\dots,k}$ can be viewed as a set $\{C_i\}_{i=1,\dots,2k}$ of second-class
constraints, for which the matrix $(\{C_i, C_j\})_{i,j=1,\dots,2k}$ is
invertible anywhere on the constraint surface.

Geometrically, this approach corresponds to defining an $(m-k)$-dimensional
submanifold of the $m$-dimensional constraint surface $\csurface$ which
intersects every gauge orbit exactly once.  It is equivalent to taking the
quotient $\cif(\csurface)/\gaugetrafos$ of functions $\cif(\csurface)$ on the
constraint surface by gauge transformations $\gaugetrafos$.\footnote{Note that
  the geometry of the constraint surface may make it impossible to choose one
  gauge orbit that globally fixes the gauge completely.  This is known as the
  Gribov obstruction \cite{Henneaux:1994aa}. In our case, however, this problem
  does not occur.}

\begin{figure}
  \centering
  \includegraphics{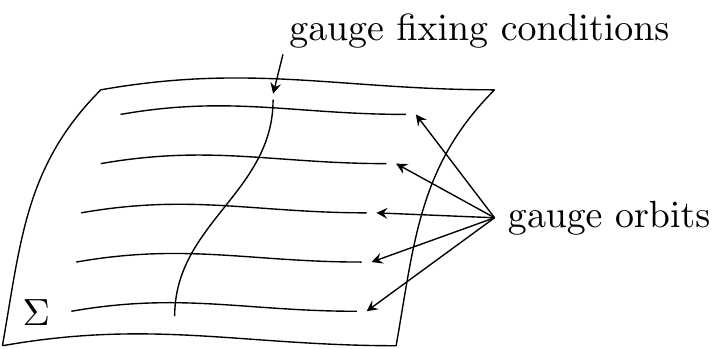}
  \caption{The constraint surface $\csurface$ inside the extended phase space
    $\phasespace$.  The gauge fixing conditions must be such that the associated submanifold of $\csurface$  intersects each gauge orbit exactly once.}
  \label{fig:constraint-surface}
\end{figure}

By adding the gauge fixing conditions, one thus replaces the original constrained system by a constrained system with second-class constraints only. In the latter,  every phase space function represents
an observable, but we need to identify functions that agree on the gauge-invariant phase space $\gaugeinvpspace \defeq \{q \in \phasespace \mid C_i(q) =
0 \ \forall i=1,\dots,2k\}$.  To ensure consistency of the constraints and gauge
fixing conditions with the Poisson structure, the Poisson structure needs to be altered in such a way
that the constraint functions and gauge fixing conditions weakly Poisson-commute
with all functions on the extended phase space and
that the new Poisson structure agrees with the original one for all
observables. Such a way of altering the Poisson structure is provided by Dirac's
gauge fixing procedure which results in the Dirac bracket \cite{Dirac:1950aa,Dirac:1951aa,Dirac:2001bh}.

To implement Dirac's gauge fixing procedure, one considers the Dirac matrix
\begin{equation}\label{dirac-matrix}
  D_{ij} \defeq \{C_i, C_j\}, \qquad i,j=1,\dots,2k.
\end{equation}
By construction it is invertible on the constraint surface $\csurface$. This implies that there
is a matrix $D^{ij}$, unique up to functions of the constraints, that satisfies
$D_{ij} D^{jk} \approx \delta_i^k$.  The Dirac bracket of functions $F, G \in
\cif(\phasespace)$ is defined in terms of this matrix as
\begin{equation}\label{dirac-bracket}
  \{F, G\}_D \defeq \{F, G\} - \sum_{i,j=1}^{2k} \{F, C_i\} D^{ij} \{C_j, G\}.
\end{equation}
It is shown in \cite{Dirac:2001bh} that the Dirac bracket defines a Poisson structure
on the constraint surface. It follows directly from formula \eqref{dirac-bracket}
that the Dirac bracket $\{O,O'\}_D$ weakly agrees with the original Poisson
bracket for all observables $O,O'\in\cif(\phasespace)$ and that the Dirac
bracket of any constraint or gauge fixing condition with a function
$F\in\cif(\phasespace)$ vanishes weakly, \ie
\begin{equation}
  \{F, C_i\}_D\approx 0 \quad\forall F\in\cif(\phasespace), \ i\in\{1,\dots,2k\}.
\end{equation}
This allows one to impose the constraints  as identities without obtaining contradictions with the Poisson structure. In other words: it ensures that the Dirac
bracket, unlike the original Poisson bracket, gives a well-defined Poisson
structure on observables, \ie on
classes of functions on $\csurface$ whose values agree on the gauge-invariant phase space $\gaugeinvpspace$.

\subsection{Gauge fixing and reference frames in (2+1)-gravity}
\label{sec:gfixingcondition}

We will now consider the gauge-invariant phase space \eqref{pnctphsp} in the Chern-Simons formulation of (2+1)-gravity as a constrained dynamical system whose extended phase space is given by
Fock and Rosly's Poisson structure on $P_3^{n+2g}$.
As discussed in Section \ref{sec:cs}, the moduli space of flat $P_3$-connections \eqref{pnctphsp} is obtained from the manifold $P_3$ by restricting the holonomies associated with the punctures to fixed $P_3$-conjugacy classes \eqref{conj}, by imposing an additional relation that mirrors the defining relation of the fundamental group $\pi_1(\surf)$ and by identifying points that are related by global conjugation with $P_3$.

The extended phase space is therefore given by the group $P_3^{n+2g}$, in  which the different components represent the
holonomies along a set of generators of the fundamental group
$\pi_1(S_{g,n})$.  The system exhibits two types of constraints.

\paragraph{Mass and spin constraints for the particles}
The first set of constraints are two constraints for each particle, which
restrict the associated holonomy $M_i$ to the conjugacy class $\mathcal
C_{\mu_i,s_i}$ defined via \eqref{conj} by its mass $\mu_i$ and spin $s_i$. In
the following, we will refer to these constraints as mass and spin
constraints for particles. In terms of the variables in Theorem \ref{thm:frtheorem}, they take
the form
\begin{equation}\label{masspinconst}
  \bp_{M_i}^2-\mu_i^2\approx 0, \qquad
  \bp_{M_i}\cdot \bj_{M_i}-\mu_is_i\approx 0 \qquad \forall i=1,\dots,n.
\end{equation}
A short calculation shows that the mass and spin constraints for the particles
are Casimir functions of Fock and Rosly's Poisson algebra. They are therefore
not associated with gauge transformations and their implementation does not pose
any difficulties.

\paragraph{Closing constraints}
The second set of constraints arises from the defining relation of the
fundamental group \eqref{pi1rel} and restricts the holonomy along the curve $c$
in Figure \ref{fig:fundamental-group} to the identity. Parametrising this
holonomy as
\begin{equation}\label{pi1const}
(\exp(p^d_C J_d), \bj_C)
    \defeq M_1^\inv\cdots M_n^\inv [A_1^\inv, B_1]\cdots[A_g^\inv, B_g],
\end{equation}
we can reformulate these constraints as
\begin{equation}\label{jpconst}
  j_C^a\approx 0,\qquad p_C^a\approx 0\qquad\forall a\in\{0,1,2\},
\end{equation}
where the three-vectors $\bp_C$, $\bj_C$ are defined as functions of the
holonomies of different particles and handles by
\begin{equation}
  \begin{aligned}
    u_C &= \exp(-p_C^aJ_a)=u_{K_g} \cdots u_{K_1} u_{M_n} \cdots u_{M_1}, \\
    \bj_C &=
      \sum_{i=1}^n \Ad(u_{M_1}^\inv \cdots u_{M_{i-1}}^\inv)\bj_\mi +
      \sum_{i=1}^g \Ad(u_{M_1}^\inv \cdots u_{M_n}^\inv u_{K_1}^\inv \cdots u_\ki^\inv)\bj_\ki,
  \end{aligned}
\end{equation}
with
\begin{equation}\label{handlej}
  u_\ki \defeq u_\bi u_\ai^\inv u_\bi^\inv u_\ai, \;
  \bj_\ki \defeq
    [\Ad(u_\bi)-\Ad(u_\bi u_\ai^\inv)]\bj_\bi -
    [\Ad(u_\bi)-\Ad(u_\ki)]\bj_\ai.
\end{equation}
In the following, we will refer to these constraints as closing constraints.
 It is is shown in \cite{Meusburger:2003aa} that the Poisson
brackets of the six closing constraints coincide with the Lie bracket of the Poincar\'e algebra:
\begin{equation}\label{constalg}
  \{p^a_C,p^b_C\}=0, \qquad
  \{j^a_C,p^b_C\}=-\tensor{\ee}{^a^b_d}\,p^d_C, \qquad
  \{j^a_C,j^b_C\}=-\tensor{\ee}{^a^b_d}\,j^d_C.
\end{equation}
The closing constraints thus form a set of first-class constraints in the
terminology of Dirac.  Their Poisson brackets with the momentum and angular
momentum three-vectors of different particles and handles take the form
\begin{equation}
  \begin{aligned}\label{gtrafos}
    \{j^a_C, p^b_X\}&=-\tensor{\ee}{^a^b_d}\,p^d_X, &\qquad
    \{p^a_C, p^b_X\}&=0, \\
    \{j^a_C, j^b_X\}&=-\tensor{\ee}{^a^b_d}\,j^d_X, &\qquad
    \{p^a_C,j^b_X\}&=\tidadi{X}{^b_c}\,T^\inv(\bp_C)^{ca},
  \end{aligned}
\end{equation}
for all $X\in\{M_1,\dots,M_n,A_1,B_1,\dots,A_g,B_g\}$.  The angular momentum
$\bj_C$ thus generates global Lorentz transformations which act on the
holonomies by simultaneous conjugation with elements of $PSL(2,\RR)$.  The momentum
$\bp_C$ generates global translations. Note that the last formula in \eqref{gtrafos} does
not correspond to the infinitesimal version of the standard expression \eqref{ttrnsf} for translations. Instead, these
translations are deformed with the invertible map $T^\inv(\bp_C)$ defined in
\eqref{tdef}.  The impact of this deformation is discussed in detail in
\cite{Meusburger:2003aa}. For the following, it will be sufficient to note that this map is
invertible and that the components of the three-vector $\bp_C$ therefore generate the
full set of translations in $\mathbb R^3$.

To summarise, the extended phase space is given by the manifold $P_3^{n+2g}$ with the Poisson structure from Theorem \ref{thm:frtheorem}.
The constraint functions are the $2n$ mass and spin
constraints \eqref{masspinconst} which act as Casimir functions,  and the six closing constraints \eqref{jpconst} which are first-class and generate six global Poincar\'e transformations.

%\subsubsection{Gauge fixing and observers}

In the geometrical picture that describes the construction of spacetimes from
regions $D\subset\mathbb M^3$ in Minkowski space, the  Poincar\'e transformations  that are generated by the first-class constraints correspond to global Poincar\'e transformations acting on $D$. As they are isometries of Minkowski space, they do not affect the
 metric on the spacetime obtained by gluing the boundary of $D$.
 Global Poincar\'e transformations therefore relate different descriptions of the same spacetime. However, they change the momentum and angular momentum variables  associated with the different particles and handles.

 As discussed in Section \ref{sec:3dgrav}, the
momentum and angular momentum variables associated with the particles and
handles characterise the geometry of the spacetime with respect to an
arbitrarily chosen reference frame, namely an observer at rest at the origin. Their change under a global Poincar\'e transformation can therefore be interpreted as a transition between two observers.

Two sets of holonomies that are related by global conjugation thus give different descriptions of the same  spacetime with respect to two different observers.
As there is no preferred observer, these two
descriptions are physically equivalent.
The two sets of holonomies therefore give equivalent descriptions of the same physical state, and the global Poincar\'e transformations relating them have the interpretation of gauge transformations.

Eliminating this gauge freedom through gauge fixing  thus amounts to choosing an observer. As there is no external or preferred reference frame, this observer must be specified with respect to the geometry of the
spacetime itself, which is given by the holonomies. This amounts to imposing a condition on certain holonomies which is not preserved under global conjugation with $P_3$. In the next section, we will give two specific preocedures in which an observer is specified either with respect to point particles in the spacetime or to the geometry of one of its handles.

\section{Gauge fixing conditions}

\label{sec:gfix}

\subsection{Gauge fixing with respect to particles}
\label{sec:partgfix}

We start by considering spacetimes $M\approx I\times S_{g,n}$ with non-trivial physical degrees of freedom ($12g+4n\geq 12$) that contain at
least two particles ($n\geq 2$) and impose a gauge fixing condition that specifies a
reference frame with respect to two particles. For simplicity, we assume that the
two particles under consideration are the ones associated with the holonomies
$M_1$ and $M_2$. As different orderings of the particles are related by the
action of the mapping class  group, which acts on Fock and Rosly's Poisson algebra
\eqref{frbracket} by Poisson isomorphisms \cite{Meusburger:2005wk}, this does not affect the
generality of the result.

We restrict attention to
spacetimes for which there are at least two particles whose associated
worldlines in Minkowski space are not parallel, \ie whose momentum
three-vectors are linearly independent. This restricts generality, as it excludes a certain sector of the gauge-invariant phase space \eqref{pnctphsp}. Note, however, that spacetimes in which all particle worldlines are parallel have a preferred direction and differ in their geometrical properties from the generic case.
We expect that they could be treated in a similar fashion. However, they will require different gauge fixing conditions and for this reason we will not consider them in this article.

To determine a gauge fixing condition, we note that a physical system consisting of two point particles in Minkowski space has eight parameters, but only two Poincar\'e-invariant degrees of freedom. The
first is the relative velocity of the particles, or, equivalently, the rapidity
of the Lorentz boost relating their momentum vectors. The second is the minimal
distance of their worldlines in Minkowski space. All other parameters  can be eliminated by applying a global Poincar\'e transformation to the system.
 A natural
condition for defining a reference frame with respect to a two-particle system is
given as follows:
\begin{enumerate}
\item One imposes that the first particle, \ie the particle associated with $M_1$, is at rest at the
  origin. This fixes the reference frame up to rotations around the $x_0$-axis
  and translations in the direction of the $x_0$-axis.
\item One imposes that the second particle, \ie the particle associated with $M_2$, moves in the direction
  of increasing $x_1$-coordinate and that the distance of its worldline from the one for $M_1$ is minimal
  at $x_0=0$. {This condition implies that the timelike, future-directed geodesic
  associated with $M_2$ intersects the $x_1$-$x_2$-plane on the $x_2$-axis.} It
  eliminates the residual gauge freedom of rotations around and translations in
  the direction of the $x_0$-axis.
\end{enumerate}
Clearly, these gauge fixing conditions define a preferred reference frame and hence an observer. The momentum rest frame of this observer coincides with the momentum rest frame of the first particle. His $x_1$-axis coincides with the direction of motion of the second particle, and his position is such that the first particle is at rest at the origin, while the second particle  is closest to the first at eigentime $t=0$.

\begin{figure}
  \centering
  \includegraphics{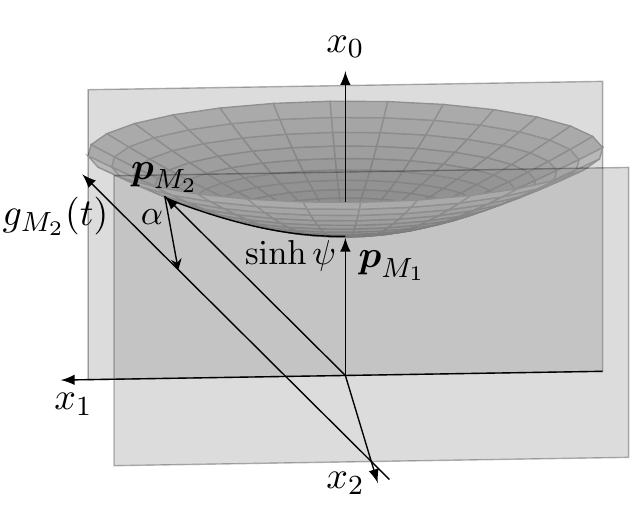}
  \caption{Gauge fixing condition for particles. The geodesic  $g_{M_1}$ associated with the first particle coincides with the $x_0$-axis, while the geodesic $g_{M_2}$ associated with the second particle lies in an affine plane parallel to the $x_0$-$x_1$plane.}
  \label{fig:gauge-fixing-conditions}
\end{figure}

These gauge fixing conditions imply that the associated timelike geodesics in
Minkowski space take the form shown  in Figure
\ref{fig:gauge-fixing-conditions}.  They can be parametrised as
\begin{equation}\label{partgeods}
  g_{M_1}(t)=(t,0,0), \qquad
  g_{M_2}(t)=(0,0,\alpha)+t\,(\cosh\psi,\sinh\psi,0), \qquad
  \alpha\in\mathbb R, \psi>0,
\end{equation}
where $\tanh\psi (0,1,0)$ is the velocity of the second particle with respect to the observer and  $\alpha (0,0,1)$ is its position at the point of minimal distance from the observer. 
Note that $\psi$ is a positive real number, since the particle moves along the $x_1$-axis in the direction of {\em increasing} $x_1$-coordinate, while $\alpha$ can take any real value, including zero.

The associated particle holonomies $M_1$, $M_2$ are determined by the condition
that they preserve the particle geodesics \eqref{partgeods} and that they lie in
a fixed $P_3$-conjugacy class specified by the masses $\mu_1$, $\mu_2$ and
the spins $s_1$, $s_2$ of the particles.  With the parametrisation
\begin{equation}
  M_i^\inv=\bigl(\exp(p^c_\mi J_c), \bj_\mi\bigr) \quad \forall i=1,2,
\end{equation}
one finds that that they are given by the following conditions on the particle's
momentum and angular momentum vectors:
\begin{subequations}\label{partpjcond}
  \begin{align}
    \bp_{M_1}&=\mu_1\dot g_{M_1}(t)=\mu_{1}\begin{pmatrix}1\\0\\0\end{pmatrix}, &
    \bj_{M_1}&=s_{1}\begin{pmatrix}1\\0\\0\end{pmatrix}, \\
    \bp_{M_2}&=\mu_2\dot g_{M_2}(t)=\mu_{2}\begin{pmatrix}\cosh\psi \\ \sinh\psi \\ 0\end{pmatrix}, &
    \bj_{M_2}&=s_{2}\begin{pmatrix}\cosh\psi \\ \sinh\psi \\ 0\end{pmatrix}
      +\alpha\begin{pmatrix}
        -\sinh\psi\sin\mu_{2} \\
        -\cosh\psi\sin\mu_{2} \\
        1-\cos\mu_{2}
      \end{pmatrix}.
  \end{align}
\end{subequations}
The  holonomies associated with the two particles are thus determined uniquely by the choice of a
reference frame together with the relative velocity $\psi$ and the minimum distance $\alpha$, and this choice eliminates the gauge freedom of applying global
Poincar\'e transformations. It can
be expressed  in terms of the following six gauge fixing conditions:
\begin{equation}\label{partconstraints}
  \begin{aligned}
    p_{M_1}^1 &\approx 0, &\quad p_{M_1}^2 &\approx 0, &\quad &p_{M_2}^2 \approx 0,\\
    j_{M_1}^1 &\approx 0, &\quad j_{M_1}^2 &\approx 0, &\quad
    &\hp_{M_2}^1 j_{M_2}^2 +
      \tan\tfrac{\mu_2}{2}\bigl(
        \hp_{M_2}^0 \hp_{M_2}^1 j_{M_2}^1 -
        \hp_{M_2}^1 \hp_{M_2}^1 j_{M_2}^0\bigr) \approx 0,
  \end{aligned}
\end{equation}
where the last and most complicated condition encodes the identity $\bx_{M_2} \approx (0, 0, \alpha)$.

In the following, it will be useful to give an explicit parametrisation of the
momentum and angular momentum vector associated with the two gauge-fixed
particles.  Parametrising the product of the associated holonomies as
\begin{equation}\label{resquants}
  \bigl(\exp(p_R^cJ_c), \bj_R\bigr) \defeq M_2M_1,
\end{equation}
we obtain an expression for the momentum three-vector
$\bp_{R}=\mu_{R}\,\hat\bp_{R}$ in terms of the dynamical parameters $\alpha$ and
$\psi$ and of the masses and spins of the gauge-fixed particles:
\begin{subequations}\label{mupres}
  \begin{align}\label{mures}
    c_{p_R}(\tfrac{\mu_{R}}{2}) &=
      \cos\tfrac{\mu_{1}}{2}\cos\tfrac{\mu_{2}}{2}
        -\sin\tfrac{\mu_{1}}{2}\sin\tfrac{\mu_{2}}{2}\cosh\psi, \\
  \label{pres}  \frac{s_{p_R}(\tfrac {\mu_R} 2)}{\mu_R}\bp_{R} &=
      -{\sin\tfrac{\mu_{1}} 2\sin\tfrac{\mu_{2}} 2}
      \begin{pmatrix}
        \cot\tfrac{\mu_2} 2 + \cot\tfrac{\mu_{1}} 2\cosh\psi \\
        \cot\tfrac{\mu_{1}} 2\sinh\psi \\
        - \sinh\psi
      \end{pmatrix},
  \end{align}
\end{subequations}
where the  functions $s_p, c_p$ are defined as in
\eqref{gencossin}.  The variable $\mu_R$ defines the opening angle of the
generalised cone associated with the two-particle system and the vector
$\hat\bp_R$ the direction of its axis.  Similarly, the angular momentum vector
$\bj_{R}$ defines the time shift of the cone, which is given by the spin
$s_{R}$, and the cone's offset $\bx_{R}$ orthogonal to its axis:
\begin{equation}\label{jres}
  \bj_{R}=s_{R}\,\hbp_{R}+[\mathds{1}-\Ad(\exp(-p_{R}^cJ_c))]\bx_{R}, \qquad
  \bx_{R}\cdot \hbp_{R}=0.
\end{equation}
In terms of the variables $\alpha,\psi$ and the masses and spins of the
gauge-fixed particles, these quantities take the form
\begin{subequations}\label{jsxres}
\begin{align}\label{jresconcrete}
  \bj_{R}&\!=\!
    s_1\begin{pmatrix} \sinh^2\psi\cos\mu_{2} -\cosh^2\psi\\-\sinh(2\psi)\sin^2\tfrac{\mu_{2}} 2\\ \sin\mu_{2}\sinh\psi\end{pmatrix}
    -s_2\begin{pmatrix} \cosh\psi\\ \sinh\psi\\ 0\end{pmatrix}
    +\alpha\begin{pmatrix}\sin\mu_{2}\sinh\psi\\ \sin\mu_{2}\cosh\psi\\ 1-\cos\mu_{2}\end{pmatrix},\\
  \allowpagebreakhere
  \label{srdef}%\raisetag{2em}
  s_{R}&\!=\!
    \frac{\sin\tfrac{\mu_{1}}2 \sin\tfrac{\mu_{2}}2}{\sin\tfrac{\mu_{r}}2}\Bigl[
      s_{1}(\cot\tfrac{\mu_{2}} 2 \!+\! \cot\tfrac{\mu_{1}}2\cosh\psi) \!+\!
      s_{2}(\cot\tfrac{\mu_{1}} 2 \!+\! \cot\tfrac{\mu_{2}}2\cosh\psi) \!-\!
      2\alpha\sinh\psi\Bigr], \\
  \allowpagebreakhere
  \begin{split}
    \bx_{R}&\!=\!
     - \frac{s_1\sin^2\tfrac{\mu_{2}}2 \sinh\psi}{2\sin^2\tfrac{\mu_{R}} 2}
         \begin{pmatrix}\sinh\psi\\\cosh\psi\\\!-\!\cot\tfrac{\mu_{2}}2\!\end{pmatrix}
     - \frac{s_2\sin^2\tfrac{\mu_{1}}2 \sinh\psi}{2\sin^2\tfrac{\mu_{R}} 2}
         \begin{pmatrix}0\\1\\\cot\tfrac{\mu_{1}} 2\end{pmatrix} \\ &\quad
      + \frac{2\alpha \sin^2\tfrac{\mu_{1}}2 \sin^2\tfrac{\mu_{2}}2 (\cot\tfrac{\mu_{1}}2+\cot\tfrac{\mu_{2}} 2\cosh\psi)}{2\sin^2\tfrac{\mu_{R}} 2}
          \begin{pmatrix}0\\1\\\cot\tfrac{\mu_{1}} 2\end{pmatrix}.
  \end{split}
\end{align}
\end{subequations}

\subsection{Gauge fixing with respect to handles}
\label{sec:handlefixing}

In addition to the particle gauge fixing conditions, we also consider two gauge
fixing conditions that specify a frame of reference with respect to the geometry of a handle.
For simplicity, we restrict attention to spacetimes $M\approx I\times S_{g,0}$ without particles and with at least two handles. The case of genus $g=1$ and no particles is the torus universe, for which the constraints can be solved explicitly (for an overview over the torus universe, see \cite{Carlip:2003aa}). The case of genus $g\geq 1$ and with particles can be treated in similar fashion, but the ordering in Figure \ref{fig:fundamental-group} should  be adjusted accordingly to simplify the calculations.

We consider two gauge fixing conditions that specify a reference frame with respect to the geometry of the first handle, which is given by the
holonomies $A_1$ and $B_1$. In this case, the Lorentzian components of the holonomies $A_1$, $B_1$ are hyperbolic. As the holonomies for the handles are not restricted to fixed $P_3$-conjugacy classes, the variables which generalise the mass- and spin variables for particles are not Casimir functions of the Fock-Rosly bracket.
Each handle has six Poincar\'e-invariant  degrees of
freedom: the mass variables
\begin{align}\label{handlemass}
\mu_{A_1}=\sqrt{|\bp_{A_1}^2|}, \qquad \quad \mu_{B_1}=\sqrt{|\bp_{B_1}^2|},
\end{align}
the spin variables defined through
\begin{align}\label{handlespin}
\mu_{A_1}s_{A_1}=\bp_{A_1}\cdot\bj_{A_1}, \qquad\quad \mu_{B_1}s_{B_1}=\bp_{B_1}\cdot\bj_{B_1},
\end{align}
as well as
 two further parameters which specify the relative orientation and
relative offset of the geodesics that are stabilised by the holonomies $A_1$ and $B_1$.

\begin{figure}
  \centering
  \includegraphics{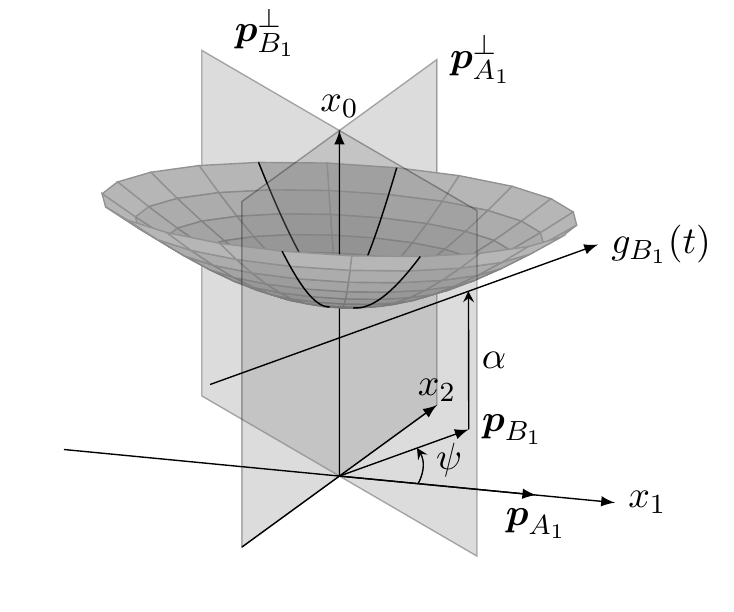}
  \includegraphics{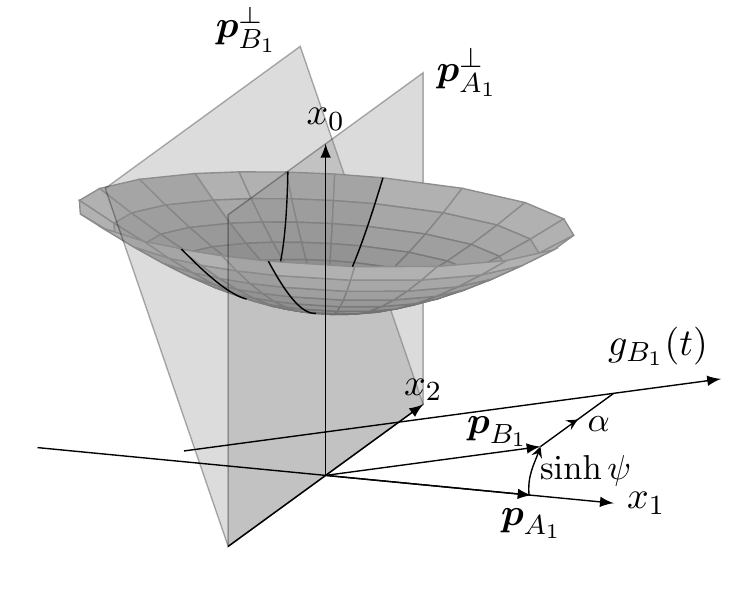}
  \caption{Gauge fixing conditions for handles.
The geodesic $g_{A_1}$ stabilised by $A_1$ coincides with the $x_1$-axis.
  For  the ``timelike intersection'' gauge fixing on the left, the geodesic
  $g_{B_1}$
  stabilised by $B_1$ lies in an affine plane parallel to the $x_1$-$x_2$ plane. The two planes  stabilised by the Lorentzian components of $A_1$ and $B_1$ intersect in the $x_0$-axis.
  For the ``spacelike intersection'' gauge fixing pictured on the right, the geodesic $g_{B_1}$ stabilised by $B_1$ lies in an affine plane parallel to the $x_0$-$x_1$-plane. The two planes  stabilised by the Lorentzian components of $A_1$ and $B_1$ intersect in the $x_2$-axis.}
  \label{fig:gauge-fixing-conditions-handle}
\end{figure}

This leads to two natural gauge fixing choices.  The first one imposes that the geodesic in Minkowski space
stabilised by $A_1$ is the $x_1$-axis, while
that geodesic stabilised by
 $B_1$ lies in an affine plane parallel to  the $x_1$-$x_2$-plane. In this case, the intersection of the two planes that are stabilised by the Lorentz components of
  the holonomies $A_1$, $B_1$ is the $x_0$ axis and hence a timelike geodesic in Minkowski space as shown in Figure \ref{fig:gauge-fixing-conditions-handle}. For this reason, we will refer to this gauge fixing  as the
``timelike intersection'' gauge fixing condition in the following.

  The second gauge fixing condition imposes again that the geodesic stabilised by  $A_1$ coincides with  the $x_1$-axis.
   However,  in this case  the geodesic stabilised by $B_1$ is required to lie in an affine plane parallel to the $x_0$-$x_1$ plane. This condition implies that the two planes that are stabilised by the Lorentzian components of
  the holonomies $A_1$, $B_1$ intersect in the $x_2$-axis as shown in Figure \ref{fig:gauge-fixing-conditions-handle}. For this reason, we will refer to this condition as the ``spacelike intersection''
gauge fixing in the following.

One might also consider imposing a ``lightlike intersection'' gauge fixing condition, in which the two planes stabilised by the holonomies $A_1,B_1$ intersect in a lightlike geodesic in Minkowski space. Although such a condition could be imposed in principle, its geometrical interpretation does not fit the case considered here, namely  spacetimes of topology $\mathbb R\times S_{g,n}$, where $S_{g,n}$ is a compact surface with punctures representing particles.  
Imposing a lightlike gauge fixing condition would imply  that the geodesics in hyperbolic space that are stabilised by the Lorentzian components of the holonomies $A_1,B_1$ meet at a point at its boundary. This situation can arise for spacetimes  whose spatial surfaces exhibit  cusps, 
but not for spacetimes with compact spatial surfaces that contain only punctures representing particles. For this reason, we restrict attention to the ``timelike intersection'' and ``spacelike intersection'' gauge fixing conditions defined above.

For the ``timelike intersection'' gauge fixing condition, the two geodesics that are stabilised  by the holonomies $A_1,B_1$ can be parametrised uniquely as
\begin{equation}\label{handle-timelike-geodesics}
  g_{A_1}(t)=(0,t,0), \qquad g_{B_1}(t)=(\alpha,0,0)+t\,(0,\cos\psi,\sin\psi),
    \qquad \alpha\in\mathbb R, \psi \in (0,\pi),
\end{equation}
which implies that the holonomies are given by
\begin{equation}\label{handle-timelike-gauge}
  \begin{aligned}
    \bp_{A_1}&=\mu_{A_1}\begin{pmatrix}0\\1\\0\end{pmatrix}, &
    \bp_{B_1}&=\mu_{B_1}\begin{pmatrix}0 \\ \cos\psi \\ \sin\psi\end{pmatrix},\\
    \bj_{A_1}&=s_{A_1}\begin{pmatrix}0\\-1\\0\end{pmatrix}, &
    \bj_{B_1}&=s_{B_1}\begin{pmatrix}0 \\ -\cos\psi \\ -\sin\psi\end{pmatrix}
      +\alpha\begin{pmatrix}
        1-\cosh\mu_{B_1} \\
        \sin\psi\sinh\mu_{B_1} \\
        -\cos\psi\sinh\mu_{B_1}
      \end{pmatrix}.
  \end{aligned}
\end{equation}
Note that we exclude the values $\psi=0,\pi$,  because they would imply that the Lorentzian components of  $A_1$ and $B_1$ stabilise the same geodesic in hyperbolic space. In this case, the holonomies $A_1$ and $B_1$ would not describe the geometry of a handle.

Equivalently, this gauge fixing choice is given by the six constraints
\begin{equation}\label{handle-timelike-constraints}
  \begin{aligned}
    p_{A_1}^0&\approx 0, &\quad p_{A_1}^2&\approx 0, &\quad &p_{B_1}^0\approx 0,\\
    j_{A_1}^0&\approx 0, &\quad j_{A_1}^2&\approx 0, &\quad
    &-\hp_{B_1}^1 j_{B_1}^0 +
      \tanh\tfrac{\mu_{B_1}}{2}\bigl(
        \hp_{B_1}^1 \hp_{B_1}^1 j_{B_1}^2 -
        \hp_{B_1}^1 \hp_{B_1}^2 j_{B_1}^1\bigr)\approx 0.
  \end{aligned}
\end{equation}

For the ``spacelike intersection'' gauge fixing condition, the two geodesics that are stabilised by the holonomies $A_1,B_1$  can be parametrised uniquely as
\begin{equation}\label{handle-spacelike-geodesics}
  g_{A_1}(t)=(0,t,0), \qquad g_{B_1}(t)=(0,0,\alpha)+t\,(\sinh\psi,\cosh\psi,0),
    \qquad \alpha\in\RR, \psi >0,
\end{equation}
and the associated holonomies take the form
\begin{equation}\label{handle-spacelike-gauge}
  \begin{aligned}
    \bp_{A_1}&=\mu_{A_1}\begin{pmatrix}0\\1\\0\end{pmatrix}, &
    \bp_{B_1}&=\mu_{B_1}\begin{pmatrix}\sinh\psi \\ \cosh\psi \\ 0\end{pmatrix},\\
    \bj_{A_1}&=s_{A_1}\begin{pmatrix}0\\-1\\0\end{pmatrix}, &
    \bj_{B_1}&=s_{B_1}\begin{pmatrix}-\sinh\psi \\ -\cosh\psi \\ 0\end{pmatrix}
      +\alpha\begin{pmatrix}
        -\cosh\psi\sinh\mu_{B_1} \\
        -\sinh\psi\sinh\mu_{B_1} \\
        1-\cosh\mu_{B_1}
      \end{pmatrix}.
  \end{aligned}
\end{equation}
As in the ``timelike intersection'' gauge fixing, the value $\psi=0$ is excluded, because it would imply that the Lorentzian components of the holonomies $A_1$, $B_1$ stabilise the same geodesic in hyperbolic space. Such a configuration would not describe the geometry of a handle.

The constraints that implement this condition can therefore be expressed as
\begin{equation}\label{handle-spacelike-constraints}
  \begin{aligned}
    p_{A_1}^0&\approx 0, &\quad p_{A_1}^2&\approx 0, &\quad &p_{B_1}^2\approx 0,\\
    j_{A_1}^0&\approx 0, &\quad j_{A_1}^2&\approx 0, &\quad
    &-\hp_{B_1}^1 j_{B_1}^2 +
      \tanh\tfrac{\mu_{B_1}}{2}\bigl(
        \hp_{B_1}^1 \hp_{B_1}^1 j_{B_1}^0 -
        \hp_{B_1}^0 \hp_{B_1}^1 j_{B_1}^1\bigr)\approx 0.
  \end{aligned}
\end{equation}
For both gauge fixing choices, the total momentum vector and angular momentum vector of the gauge-fixed  handle are defined via the identity
\begin{align}\label{pkhandle}
&(\exp(p^a_RJ_a), \bj_R)=[B_1,A_1^\inv]\\
&\quad=\left([u_{B_1}, u_{A_1}^\inv]\;,\; -\left[\Ad(u_{B_1})-\Ad(u_{B_1}u_{A_1}^\inv)\right]\bj_{B_1}+\left[\Ad(u_{B_1})-\Ad([u_{B_1}, u_{A_1}^\inv])\right]\bj_{A_1}\right).\nonumber
\end{align}
%\begin{align}
%  A_1^\inv=\bigl(\exp(p^c_{A_1} J_c), \bj_{A_1}\bigr),\quad
%  B_1^\inv=\bigl(\exp(p^c_{B_1} J_c), \bj_{B_1}\bigr)
%\end{align}
%
%\begin{equation}
%  [B_1, A_1^\inv]^\inv = A_1^\inv B_1 A_1 B_1^\inv \eqdef (\exp(p_{K_1}^cJ_c), \bj_{K_1}),
%\end{equation}
%\begin{equation}\label{handle-spacelike-pK}
%  \begin{aligned}
%    \cosh\tfrac{\mu_{K_1}}{2}&=\cosh^2\tfrac{\mu_{A_1}}{2} + \sinh^2\tfrac{\mu_{A_1}}{2}\Bigl( \cosh\mu_{B_1}\sinh^2\psi - \cosh^2\psi \Bigr), \\
%    \hat\bp_{K_1} &= \frac{2\sinh^2\tfrac{\mu_{A_1}} 2\sinh^2\tfrac{\mu_{B_1}} 2\sinh\psi}{\sinh\tfrac{\mu_{K_1}} 2}
%      \begin{pmatrix}
%        \coth\tfrac{\mu_{A_1}} 2\cosh\psi-\coth\tfrac{\mu_{B_1}} 2 \\
%        \sinh\psi\coth\tfrac{\mu_{A_1}} 2 \\
%        \cosh\psi-\coth\tfrac{\mu_{A_1}} 2 \coth\tfrac{\mu_{B_1}} 2
%      \end{pmatrix}.
%  \end{aligned}
%\end{equation}
%
%\begin{equation}
%  \begin{split}
%    \bj_{K_1} = \text{very ugly},
%  \end{split}
%\end{equation}
%\begin{equation}\label{handle-spacelike-sK}
%  s_{K_1} = \frac{2 \sinh^2\tfrac{\mu_{A_1}}{2} \sinh^2\tfrac{\mu_{B_1}}{2} \sinh\psi}{\sinh\tfrac{\mu_{K_1}}{2}} \Bigl[
%      4 \alpha \cosh\psi
%    + 2 s_{A_1} \coth\tfrac{\mu_{A_1}}{2} \sinh\psi
%    + 2 s_{B_1} \coth\tfrac{\mu_{B_1}}{2} \sinh\psi\Bigr]
%\end{equation}
The geometrical interpretation of these gauge fixing conditions is less immediate than in the particle case. They specify the observer with respect to the geometry of the spacetime, namely the holonomy variables $A_1,B_1$ which characterise the geometry of its first handle.
Similar conditions for fixing an observer are investigated in  \cite{Meusburger:2009aa}, which studies measurements performed by  observers by emitting and receiving returning lightrays. It is shown there that the measurements of  observers specified by such gauge fixing conditions take a particularly simple form and that the conditions can be interpreted as defining an observer who is  ``comoving'' with respect to certain geodesics in the spacetime.

%%%%%%%%%%%%%%%%%%%%%%%%%%%%%%%%%%%

\section{The Dirac bracket for (2+1)-gravity}
\label{sec:results}

\subsection{The Dirac bracket for the particle gauge fixing condition}
\label{diracpart}

We are now ready to derive the central result of our paper, which is an
explicit description of the Dirac bracket associated with the
gauge fixing conditions for particles and handles. We start by considering the gauge fixing based on two particle holonomies  with the
 first-class
constraints \eqref{jpconst} and the gauge fixing conditions
\eqref{partconstraints}.  We have the following theorem.

\begin{theorem}[Dirac bracket for the particle gauge fixing]\label{thm:particledirac}\mbox{}

In terms of the variables defined in Theorem \ref{thm:frtheorem} and in terms of
the parametrisation \eqref{partpjcond} by the variables $\psi,\alpha$ defined in
\eqref{partgeods}, the Dirac bracket resulting from the gauge fixing conditions
\eqref{partconstraints} is given as follows.
\begin{enumerate}
\item The components of the vectors $\bp_{M_1}$, $\bj_{M_1}$ are
  Casimir functions for the Dirac bracket.

\item The Dirac brackets of $\bp_{M_2}, \bj_{M_2}$ with $\bp_{M_2}, \bj_{M_2}$
  vanish: $\{\psi,\alpha\}_D \approx 0$.  The Dirac brackets of $\bp_{M_2},
  \bj_{M_2}$ with momenta and angular momenta $\bp_X,\bj_X$,
  $X\in\{M_3,\dots,M_n,\allowlinebreakhere A_1,\dots,B_g\}$, are given by the
  following brackets:
  \begin{gather}
    \{\psi, \bp_X\}_D \approx 0, \quad
    \{\psi, \bj_X \}_D \!\approx\!
      -\frac{\idadi{X} \hbp_{R}}{\hbp_R \cdot (\partial \bj_R / \partial \alpha)}, \quad
    \{\alpha, \bp_X\}_D \!\approx\!
      \frac{ \hbp_{R} \wedge \bp_X}{\hbp_R \cdot (\partial \bj_R / \partial \alpha)}, \nonumber \\
    \{\alpha, \bj_X\}_D \!\approx\!
        \frac{\hbp_{R}\wedge \bj_X}{\hbp_R \cdot (\partial \bj_R / \partial \alpha)}
          \! -\! \frac{  \idadi{X}\bigr[   \hbp_{R} \wedge \bx_{R}
          \!-\! \xi_R\,
            \, \hbp_{R}\bigr]}{\hbp_R \cdot (\partial \bj_R / \partial \alpha)},
    \label{p2brack}
  \end{gather}
  where $\xi_R= \frac{2\, \be_2 \cdot (\hbp_{R} \wedge \bx_{R})}{\hbp_R \cdot (\partial \bj_R / \partial \alpha)}$.

\item For all $X,Y\in\{M_3,\dots,M_n, A_1,\dots,B_g\}$ the Dirac brackets of the
  associated momenta and angular momenta take the form
  \begin{subequations}\label{dirac-non-gauge-fixed}
    \begin{align}
      \label{ppdirac}
      \{p^a_X,p^b_Y\}_D &\approx 0,  \\
      \label{pjdirac}
      \{j^a_X,p^b_Y\}_D &\approx
        \{j^a_X,p^b_Y\}+
        \tidadi{X}{^a_c}\tensor{V}{^c_d}\tensor{\ee}{^{db}_f}\ p^f_Y, \\
      \label{jjdirac}\raisetag{1.3em}
      \begin{split}
        \{j^a_X,j^b_Y\}_D &\approx
          \{j^a_X,j^b_Y\}+
          \tidadi{X}{^a_g}\tensor{V}{^g_d}\tensor{\ee}{^{db}_f}j^f_Y -
          \tidadi{Y}{^b_g}\tensor{V}{^g_d}\tensor{\ee}{^{da}_f}j^f_X \\
          &\qquad\qquad\quad+\tidadi{X}{^{ac}}\tidadi{Y}{^{bd}}\ee_{cdf} m^f,
      \end{split}
    \end{align}
  \end{subequations}
  with
  \begin{equation}\label{V-definition}
    V_{ab} = \tfrac 1 2 \eta_{ab}+\tfrac 1 2 \ee_{abc}w^c,
  \end{equation}
  and
  \begin{equation}\label{wmdef}
    \bw=
      \begin{pmatrix}
        \cot\tfrac{\mu_{1}}{2} \\
        \cot\tfrac{\mu_{1}}{2}\coth\psi \\
        -\coth\psi
      \end{pmatrix}, \qquad
    \bm=%\frac{s_1}{4\sin^2\tfrac{\mu_1}{2}}\begin{pmatrix}1 \\ \coth\psi \\ 0\end{pmatrix}
    \frac{s_1(\be_2\wedge\hat\bp_{M_2})}{4\sin^2\tfrac{\mu_1} 2 \sinh\psi} +
      \frac{\alpha}{2}\frac{\partial \bw}{\partial\psi}.
  \end{equation}
\end{enumerate}
\end{theorem}

\begin{proof}

To verify these relations for the Dirac bracket we need to calculate the Dirac
matrix and invert it on the constraint surface.  As the number of constraints makes a brute-force calculation very cumbersome, we exploit the structural features of Fock and Rosly's Poisson bracket to derive
the result in a more direct way.  Firstly, we recall that the
variables $p_X^a$,
 $j_X^a$, $X\in\{M_1,...,M_n,A_1,...,B_g\}$,
can be identified, respectively,  with functions and  vector fields on $PSL(2,\RR)^{n+2g}$.
This suggest grouping  the constraints and gauge fixing conditions into two sets depending on whether they contain angular momentum variables $j^a_X$ or not. We order our constraints
 correspondingly:
  \begin{align}\label{ppjj-constraints}
    C_1 &= j_C^0, &\quad C_2 &= j_C^1, &\quad C_3 &= j_C^2, \\
    C_4 &= j_{M_1}^1, &\quad C_5 &= j_{M_1}^2, &\quad
    C_6 &= \hp_{M_2}^1 j_{M_2}^2 +
      \tan\tfrac{\mu_2}{2}\bigl(
        \hp_{M_2}^0 \hp_{M_2}^1 j_{M_2}^1 -
        \hp_{M_2}^1 \hp_{M_2}^1 j_{M_2}^0\bigr), \nonumber\\
    C_7 &= p_C^0, &\quad C_8 &= p_C^1, &\quad C_9 &= p_C^2, \nonumber\\
    C_{10} &= p_{M_1}^1, &\quad C_{11} &= p_{M_1}^2, &\quad C_{12} &= p_{M_2}^2.\nonumber
  \end{align}
With this ordering of the constraints, the Dirac matrix takes the form
\begin{equation}
  D = \begin{pmatrix}J & P \\ -P^T & 0\end{pmatrix}, \qquad
  J \defeq (\{C_i, C_j\})_{i,j=1,\ldots,6}, \qquad
  P \defeq (\{C_i,C_{j+6}\})_{i,j=1,\ldots,6}.
\end{equation}
As suggested by the notation, the $6\times 6$-matrix $J$ contains the brackets
which involve two angular momenta. Its entries are therefore linear combinations
of angular momenta with momentum-dependent coefficients. The $6\times 6$-matrix
$P$ contains the brackets of momenta with angular momenta.  Its entries
therefore depend only on the momenta.  On the constraint surface, the matrices
$J$ and $P$ take the form
\begin{equation}
  J \approx \begin{pmatrix}0 & H \\ -H^T & G\end{pmatrix}, \qquad
  P \approx \begin{pmatrix}0 & A \\ B & C\end{pmatrix},
\end{equation}
with $3\times 3$-matrices $A, B, C, G, H$ given by
\begin{equation}\label{abchg-def}
  \left.
  \begin{gathered}
    A_{ij} \defeq \{C_i, C_{j+9}\}, \quad
    B_{ij} \defeq \{C_{i+3}, C_{j+6}\}, \quad
    C_{ij} \defeq \{C_{i+3}, C_{j+9}\}, \\
    G_{ij} \defeq \{C_{i+3}, C_{j+3}\}, \quad
    H_{ij} \defeq \{C_i, C_{j+3}\},
  \end{gathered}
  \quad\right\}\quad i,j=1,2,3.
\end{equation}
As the lower-right block of $D$ and the upper-left blocks of $J$ and $P$ vanish there,
we obtain
\begin{gather}
  D^\inv \approx \begin{pmatrix}0 & -(P^\inv)^T \\ P^\inv & P^\inv J (P^\inv)^T\end{pmatrix},\qquad\qquad
  P^\inv \approx \begin{pmatrix}-B^\inv C A^\inv & B^\inv \\ A^\inv & 0\end{pmatrix}, \\
  P^\inv J (P^\inv)^T \approx
      \begin{pmatrix}
        B^\inv\bigl[G-CA^\inv H+(CA^\inv H)^T\bigr](B^\inv)^T & -B^\inv H^T(A^\inv)^T \\
        A^\inv H (B^\inv)^T & 0
      \end{pmatrix}.\nonumber
\end{gather}
The task of inverting the Dirac matrix on the constraint surface thus reduces to
inverting the $3\times 3$-matrices $A$ and $B$, and for all $X, Y \in
\{M_1, \dots, B_g\}$ the Dirac bracket takes the form
\begin{subequations}
\begin{align}
  \{p_X^a, p_Y^b\}_D &\approx 0, \\
  \label{jp-dirac-general}\raisetag{1.7em}
  \begin{split}
    \{j_X^a, p_Y^b\}_D &\approx \{j_X^a, p_Y^b\} + \smashoperator{\sum_{i,j=1}^3} \Bigl[
      \{j_X^a, C_{i+6}\} (B^\inv)_{ij} \{p_Y^b, C_{j+3}\}
      +\{j_X^a, C_{i+9}\} (A^\inv)_{ij} \{p_Y^b, C_j\} \\
      &\hspace{8em} -\{j_X^a, C_{i+6}\} (B^\inv C A^\inv)_{ij} \{p_Y^b, C_j\}
    \Bigr],
  \end{split} \\
  \label{jj-dirac-general}\raisetag{2.2em}
  \begin{split}
    \{j^a_X, j^b_Y\}_D &\approx \{j^a_X,j^b_Y\}
                    + \sum_{i=1}^6\sum_{j=7}^{12} \{j^a_X, C_i\} (D^\inv)_{ij} \{j^b_Y, C_j\} + \sum_{i=7}^{12}\sum_{j=1}^6 \{j^a_X, C_i\} (D^\inv)_{ij} \{j^b_Y, C_j\} \\
      &\hspace{5em}  + \sum_{i=7}^{12}\sum_{j=7}^{12} \{j^a_X, C_i\} (D^\inv)_{ij} \{j^b_Y, C_j\}.
  \end{split}
\end{align}
\end{subequations}
We are now ready to prove the specific claims:
\begin{enumerate}

\item That the components $p_{M_1}^1$, $p_{M_1}^2$, $j_{M_1}^1$ and $j_{M_1}^2$
  are Casimir functions of the Dirac bracket follows directly because
  they are gauge fixing conditions.  The components $p_{M_1}^0 \approx \mu_1$ and $j_{M_1}^0
  \approx s_1$ have this property since $\mu_1$ and $s_1$ are Casimir
  functions of the Fock-Rosly Poisson structure.

\item The first equation in \eqref{p2brack} follows because  the parameter $\psi$ is
  determined by $\bp_{M_2}$. To obtain the bracket $\{\psi, \bj_X\}_D$, we
  need to compute $\{p_{M_2}^0, \bj_X\}_D$ from \eqref{jp-dirac-general} and
  then use the chain rule.  For this, we note that we have $\bp_C \approx
  0$ and hence $T^\inv(\bp_C) \approx \mathds{1}$ which implies
  \begin{equation}\label{jXC789-bracket}
    \{j_X^a, C_{i+6}\} \approx -\tidadi{X}{^a^{,i-1}} \quad \forall i = 1, 2, 3.
  \end{equation}
  Note that the index $i$ plays the role of a label on the left hand side of
  \eqref{jXC789-bracket}, while it is interpreted as a Lorentz index on the
  right hand side.  Using \eqref{jXC789-bracket}, we obtain
  \begin{equation}
    \{\psi, j_X^a\}_D
      \approx \frac{\sin\tfrac{\mu_R}{2}  \tidadi{X}{^a_b} \, \hp_R^b}{2\sin\tfrac{\mu_1}{2}\sin\tfrac{\mu_2}{2}\sinh\psi}
     = -\frac{ \tidadi{X}{^a_b} \, \hp_R^b}{\hbp_R \cdot (\partial \bj_R / \partial \alpha)}.
  \end{equation}
  The relations for $\{\alpha, \bp_X\}_D$ and $\{\alpha, \bj_X\}_D$ can be
  verified in a similar fashion. The bracket $\{\alpha,\psi\}_D$ can be
  derived analogously from the bracket $\{j_{M_2}^a, p_{M_2}^b\}_D$.

\item Relation \eqref{ppdirac} follows directly from the form of the matrix $D^\inv$.  To
  prove equation \eqref{pjdirac}, we note that for $X \in \{M_3, \dots, B_g\}$
  we have $\{j^a_X, C_{i+9}\} = 0$.  Using again \eqref{jXC789-bracket} and the
  specific form of the constraints $C_4$, $C_5$ and $C_6$, we can simplify
  expression \eqref{jp-dirac-general} to
  \begin{gather}\label{jXpY-with-V}
    \{j_X^a, p_Y^b\}_D \approx \{j_X^a, p_Y^b\}+ \tidadi{X}{^a^e} \, V_{ed} \, \tensor{\ee}{^d^b_c} \, p_Y^c, \\
    V_{ed}=
          (B^\inv C A^\inv)_{e+1,d+1}
          - \sum_{j=1}^3 (B^\inv)_{e+1,j} \smashoperator{\sum_{X \in \{M_1, M_2\}}}
              \theta_{j,a}^X \tidad{X}{_d^a} \quad \forall e,d=0,1,2,\nonumber
  \end{gather}
  where the coefficients $\theta_{j,a}^X$ are defined by the equation
  \begin{equation}
  C_{j+3}=\smashoperator{\sum_{X\in\{M_1,M_2\}}} \theta^X_{j,a}\,j^a_X \quad \forall j=1,2,3,
  \end{equation}
  and can be read off from \eqref{ppjj-constraints}.  Inserting the expressions
  for the matrices $A, B, C$ and the parameters $\theta^X_{j,a}$, we obtain
%  \begin{equation}
%    \begin{split}
%     V= (V_{ed}) &\approx
%         \frac{1}{2}\begin{pmatrix}
%           1 & -\coth\psi & -\coth\psi\cot\tfrac{\mu_1}{2} \\
%           \coth\psi & -1 & \cot\tfrac{\mu_1}{2} \\
%           \coth\psi\cot\tfrac{\mu_1}{2} & -\cot\tfrac{\mu_1}{2} & -1
%         \end{pmatrix},
%    \end{split}
%  \end{equation}
  expressions \eqref{V-definition} and \eqref{wmdef}.

\item To show relation \eqref{jjdirac} for the Dirac bracket of two angular
  momenta, we make use of equation \eqref{jXC789-bracket} and of the identity
  $\{j^a_X, C_{i+9}\}=0$ for all $X \in \{M_3, \dots, B_g\}$. Moreover, we note
  that for $i=1,\dots,6$ and $X \in \{M_3, \dots, B_g\}$ we have
  \begin{equation}
    \{p^a_X, C_i\} = f^a_{b,i} \, p_X^b, \qquad
    \{j^a_X, C_i\} = f^a_{b,i} \, j_X^b,
  \end{equation}
  with coefficient functions $f^a_{b,i} \in \cif(PSL(2,\RR)^{2n+g})$.  We can
  therefore use the results that led to \eqref{jXpY-with-V} to simplify equation
  \eqref{jj-dirac-general}:
  \begin{align}
   & \{j^a_X, j^b_Y\}_D \approx \{j^a_X,j^b_Y\}
      + \idadi{X}^{ad} \, V_{dg} \, \tensor{\ee}{^g^b_f} j_Y^f \\
      &\quad - \idadi{Y}^{bd} \, V_{dg} \, \tensor{\ee}{^g^a_f} j_X^f + \idadi{X}^{ac} \idadi{Y}^{bd} (D^\inv)_{c+7,d+7}.\nonumber
  \end{align}
  As the matrix $G$ and, consequently, the matrix $(D^\inv)_{c+7,d+7}$ is
  anti-symmetric, it is of the form $(D^\inv)_{c+7,d+7} = \tensor{\ee}{_c_d_f}
  m^f$ with a three-vector $\bm$ that depends only on $\alpha,\psi$ and on the
  parameters $\mu_1,\mu_2,s_1,s_2$.  Using the explicit expression for the matrix $D^\inv$, we find that $\bm$ takes the form \eqref{wmdef}.
\end{enumerate}

\end{proof}

The Dirac brackets derived in Theorem \ref{thm:particledirac} have a direct physical interpretation.
The gauge fixing condition \eqref{partconstraints} restricts the motion of the particle associated with the holonomy $M_1$ completely. As the mass and spin variables $\mu_{1}, s_{1}$ are external parameters, \ie Casimir functions of the Poisson bracket, there are no longer any physical degrees of freedom associated with this particle. This is reflected in the fact that the Dirac brackets of the variables $\bp_{M_1}$, $\bj_{M_1}$ with all other variables vanish.

In contrast, the  motion of the particle associated with the holonomy $M_2$ is not determined completely by the gauge fixing condition \eqref{partconstraints}. It is characterised by  two residual physical degrees of freedom. The first is its relative velocity with respect to the first particle, which is given by the parameter $\psi$. The second is its minimal distance from the first particle, which is encoded in the parameter $\alpha$. It follows from \eqref{p2brack} that the parameter $\psi$ generates via the Dirac bracket a global translation \eqref{ttrnsf} in the direction of $\hat\bp_R$  that acts on all non-gauge-fixed holonomies.
Similarly, the parameter $\alpha$ generates via the Dirac bracket a combination of a rotation \eqref{ltrnsf} around $\hat\bp_R$ and a translation \eqref{ttrnsf} in the direction of $\hat\bp_R$.

The Poisson brackets of the momenta and angular momenta  $\bp_X,\bj_X$, $X\in\{M_3$, \dots, $M_n$, $A_1$, \dots, $B_g\}$ of the non-gauge-fixed variables are modified with terms involving a matrix $V$ and the vector $\bm$  defined in \eqref{V-definition}, \eqref{wmdef}. As in the case of the Fock-Rosly Poisson structure, the Dirac brackets of two momenta vanish. Moreover, the matrix $V$, which appears in the bracket of momenta with angular momenta, depends only on the external parameters $\mu_{1}, \mu_2$ and the dynamical parameter $\psi$. It is a function of the Lorentzian components of the holonomies only. This implies that we can again identify the momenta of the non-gauge-fixed holonomies with functions on the product of $n+2g-2$ copies of the Lorentz group and define the parameter $\psi$ as a function on $PSL(2,\RR)^{n+2g-2}$ through \eqref{mupres}.

However, the Poisson brackets of two angular momentum variables are no longer linear combinations of angular momentum variables with momentum-dependent coefficients.
This is due to the fact that the vector $\bm$ defined in \eqref{wmdef} has a component which
depends on the parameters $\mu_1,s_1$ and $\psi$ only.  In contrast to the Fock-Rosly bracket, the angular momentum variables of the non-gauge-fixed holonomies can therefore not be identified with vector fields on the manifold $PSL(2,\RR)^{n+2g-2}$. Instead, they are given as a sum of such vector fields and of functions on $PSL(2,\RR)^{n+2g-2}$. We will show in Section \ref{sec:particleinterpretation} that the Dirac bracket on the constraint surface can be obtained from the Fock-Rosly bracket on $P_3^{n+2g-2}$ via a global translation.

\subsection{The Dirac bracket for the handle gauge fixing conditions}
\label{diracparthandle}

For  comparison with the gauge fixing condition based on the particle holonomies and to treat the case without particles, we will now derive the Dirac bracket for the two gauge fixing conditions \eqref{handle-timelike-constraints} and \eqref{handle-spacelike-constraints} associated with the handles.
The corresponding Dirac bracket is obtained along the same lines as the one for the particle gauge fixing condition in Theorem \ref{thm:particledirac}.
The only difference is the concrete form of the matrices $A,B,C,D,G,H$ in the proof of Theorem \ref{thm:particledirac} and the fact that the mass and spin variables for the $a$- and $b$-cycle of the gauge-fixed handle are dynamical variables instead of external parameters. After a direct calculation repeating the steps in the proof of Theorem \ref{thm:particledirac}, we obtain the following theorem.

\begin{theorem}[Dirac bracket for the handle gauge fixing]\label{thm:handledirac}\mbox{}

1. In terms of the variables defined in Theorem \ref{thm:frtheorem} and in equation
\eqref{handle-timelike-gauge}, the Dirac bracket resulting from the ``timelike intersection'' gauge fixing conditions
\eqref{handle-timelike-constraints} is given as follows.
\begin{enumerate}[a)]
\item The brackets of the mass and spin variables $\mu_{A_1}, \mu_{B_1}, s_{A_1}, s_{B_1}$ of the gauge-fixed handle and the variables $\psi,\alpha$ take the form
\begin{align}\label{tlikems}
  &\{\mu_{A_1}, s_{A_1}\!\}_D\!\approx\!
  \{\mu_{B_1}, s_{B_1}\!\}_D\! \approx\!
  \{\mu_{A_1}, \mu_{B_1}\!\}_D \!\approx\!  %&
  \{\psi, \mu_{A_1}\!\}_D \!\approx\! \{\psi, \mu_{B_1}\!\}_D\!\approx\!   \{\alpha, \psi\}_D\!\approx\! 0,\\
 &\!\!\begin{array}{ll}
  \{\mu_{A_1}, s_{B_1}\}_D \approx \{s_{A_1}, \mu_{B_1}\}_D \approx \cos\psi, &
    %&
    \{s_{A_1}, s_{B_1}\}_D \approx -\alpha\sin\psi, \\[+.3em]
    \{\psi, s_{A_1}\}_D \!\approx\! \tfrac12 \sin\psi\coth\tfrac{\mu_{B_1}}{2}, %&
  & \{\psi, s_{B_1}\}_D \!\approx\! -\tfrac12 \sin\psi\coth\tfrac{\mu_{A_1}}{2}, \\[+.3em]
  \{\alpha, s_{A_1}\}_D \!\approx\! \tfrac\alpha 2 \cos\psi\coth\tfrac{\mu_{B_1}}{2} \!-\! \tfrac {s_{B_1}}4  \sin\psi\,(\sin\tfrac{\mu_{B_1}}{2})^{-2},  &  \{\alpha, \mu_{A_1}\}_D\! \approx\! \tfrac12 \sin\psi\coth\tfrac{\mu_{B_1}}{2},\\[+.3em]
  \{\alpha, s_{B_1}\}_D \!\approx\! -\tfrac\alpha 2 \cos\psi\coth\tfrac{\mu_{A_1}}{2}\! +\! \tfrac{s_{A_1}}4  \sin\psi\,(\sin\tfrac{\mu_{A_1}}{2})^{-2}, & \{\alpha, \mu_{B_1}\}_D \!\approx\! -\tfrac12 \sin\psi\coth\tfrac{\mu_{A_1}}{2}.
  \end{array}\nonumber
\end{align}

\smallskip
\item The Dirac brackets of the  variables $\mu_{A_1}, s_{A_1}, \mu_{B_1}, s_{B_1}, \alpha,\psi$ with the momentum and angular momentum variables $p^a_X,j^a_X$  ($X\neq\{A_1,B_1\}$)  of the remaining holonomies are given by
\begin{align}\label{tlikeas}
  &\{\psi, p_X^a\}_D \approx 0, &
  &\{\psi, j_X^a\}_D \approx \tidadi{X}{^a_b} q_\infty^b, \\
  &\{\alpha, p_X^a\}_D \approx -\tensor{\ee}{^a_b_c} q_\infty^b p_X^c, &
  &\{\alpha, j_X^a\}_D \approx -\tensor{\ee}{^a_b_c} q_\infty^b j_X^c + \tidadi{X}{^a_b} r_\infty^b, \nonumber\\
  &\{\mu_{A_1}, p_X^a\}_D \approx 0, &
  &\{\mu_{A_1}, j_X^a\}_D \approx \tidadi{X}{^a_b} q_{A_1}^b, \nonumber\\
  &\{s_{A_1}, p_X^a\}_D \approx -\tensor{\ee}{^a_b_c} q_{A_1}^b p_X^c, &
  &\{s_{A_1}, j_X^a\}_D \approx -\tensor{\ee}{^a_b_c} q_{A_1}^b j_X^c, \nonumber\\
  &\{\mu_{B_1}, p_X^a\}_D \approx 0, &
  &\{\mu_{B_1}, j_X^a\}_D \approx \tidadi{X}{^a_b} q_{B_1}^b,\nonumber \\
  &\{s_{B_1}, p_X^a\}_D \approx -\tensor{\ee}{^a_b_c} q^b p_X^c, &
  &\{s_{B_1}, j_X^a\}_D \approx  - \tensor{\ee}{^a_b_c} q_{B_1}^b j_X^c+\tidadi{X}{^a_b} r_{B_1}^b,\nonumber
\end{align}
where:
\begin{align}\label{helpvecstl}
  &\bq_{A_1} \equiv \frac{\be_1}{2}, \;\;\;
  \bq_{B_1} \equiv \frac12\begin{pmatrix}
    \sin\psi\coth\tfrac{\mu_{A_1}}{2} \\
    \cos\psi \\
    \sin\psi
  \end{pmatrix},\;\;\; \bq_\infty \equiv -\frac14 \begin{pmatrix}
    1-\cos\psi\coth\tfrac{\mu_{A_1}}{2}\coth\tfrac{\mu_{B_1}}{2} \\
    \sin\psi\coth\tfrac{\mu_{B_1}}{2} \\
    \coth\tfrac{\mu_{A_1}}{2}-\cos\psi\coth\tfrac{\mu_{B_1}}{2}
  \end{pmatrix},\nonumber\\
    &\br_\infty \equiv \frac{s_{A_1}\cos\psi}{8\sinh^2\tfrac{\mu_{A_1}}{2}} \begin{pmatrix}
      -\coth\tfrac{\mu_{B_1}}{2} \\
      0 \\
      1/\cos\psi
    \end{pmatrix}\! -\!
    \frac{s_{B_1}\cos\psi}{8\sinh^2\tfrac{\mu_{B_1}}{2}} \begin{pmatrix}
      \coth\tfrac{\mu_{A_1}}{2} \\
      -\tan\psi \\
      1
    \end{pmatrix} \! -\!
    \frac{\alpha\sin\psi}{4\tanh\tfrac{\mu_{B_1}}{2}} \begin{pmatrix}
      \coth\tfrac{\mu_{A_1}}{2} \\
      \cot\psi \\
      1
    \end{pmatrix},\nonumber\\
 &\br_{B_1} \equiv \frac{\alpha}{2}\begin{pmatrix}
    \cos\psi\coth\tfrac{\mu_{A_1}}{2} \\
    -\sin\psi \\
    \cos\psi
  \end{pmatrix} -
  \frac{s_{A_1}\sin\psi\,\be_0}{4\sinh^2\tfrac{\mu_{A_1}}{2}}.
\end{align}

\smallskip
\item The Poisson brackets of the momentum and angular momentum variables $p^a_X,j^a_X$ of the non-gauge-fixed holonomies are  analogous to the ones for the particle gauge fixing. They are given by  \eqref{ppdirac}, \eqref{pjdirac}, \eqref{jjdirac} and \eqref{V-definition} with
\begin{equation}\label{handle-timelike-wmdef}
  \bw=-\frac{1}{2\sin\psi}\begin{pmatrix}
      \coth\tfrac{\mu_{A_1}}{2}\coth\tfrac{\mu_{B_1}}{2} - \cos\psi \\
      2\coth\tfrac{\mu_{A_1}}{2}\sin\psi \\
      \coth\tfrac{\mu_{B_1}}{2} - \coth\tfrac{\mu_{A_1}}{2}\cos\psi
    \end{pmatrix}, \;\;
  \bm=\frac \alpha 2 \frac{\partial \bw}{\partial\psi}+\frac {s_A} 2 \frac{\partial \bw}{\partial\mu_A}+\frac {s_B} 2 \frac{\partial \bw}{\partial\mu_B}.
   % \bm&=
    %  \frac{s_{A_1}}{8\sin\psi\sinh^2\tfrac{\mu_{A_1}}{2}} \begin{pmatrix}
    %      \coth\tfrac{\mu_{B_1}}{2} \\
    %      2\sin\psi \\
    %      -\cos\psi
    %    \end{pmatrix} +
    %  \frac{s_{B_1}}{8\sin\psi\sinh^2\tfrac{\mu_{B_1}}{2}} \begin{pmatrix}
     %     \coth\tfrac{\mu_{A_1}}{2} \\
     %     0 \\
     %     1
     %   \end{pmatrix} +
     % \frac{\alpha}{2} \frac{\partial \bw}{\partial\psi}.
\end{equation}
\end{enumerate}

\bigskip
2. The Dirac bracket resulting from the ``spacelike intersection'' gauge fixing conditions
\eqref{handle-spacelike-constraints} is given as follows.
\begin{enumerate}[a)]
\item The brackets of the mass and spin variables $\mu_{A_1}, \mu_{B_1}, s_{A_1}, s_{B_1}$ of the gauge-fixed handle and the variables $\psi,\alpha$  take a form analogous to  \eqref{tlikems}:
\begin{align}\label{slikeems}
  &\{\mu_{A_1}, s_{A_1}\!\}_D\!\approx\!
  \{\mu_{B_1}, s_{B_1}\!\}_D\! \approx\!
  \{\mu_{A_1}, \mu_{B_1}\!\}_D \!\approx\!  %&
  \{\psi, \mu_{A_1}\!\}_D \!\approx\! \{\psi, \mu_{B_1}\!\}_D\!\approx\!   \{\alpha, \psi\}_D\!\approx\! 0,\\
 &\!\!\begin{array}{ll}
  \{\mu_{A_1}, s_{B_1}\}_D \approx \{s_{A_1}, \mu_{B_1}\}_D \approx \cosh\psi, &
    %&
    \{s_{A_1}, s_{B_1}\}_D \approx \alpha\sinh\psi, \\[+.3em]
    \{\psi, s_{A_1}\}_D \!\approx\! \tfrac12 \sinh\psi\coth\tfrac{\mu_{B_1}}{2}, %&
  & \{\psi, s_{B_1}\}_D \!\approx\! -\tfrac12 \sinh\psi\coth\tfrac{\mu_{A_1}}{2}, \\[+.3em]
  \{\alpha, s_{A_1}\}_D \!\approx\! \tfrac\alpha 2 \cosh\psi\coth\tfrac{\mu_{B_1}}{2} \!-\! \tfrac{s_{B_1}}4 \sinh\psi\,(\sin\tfrac{\mu_{B_1}}{2})^{-2},  &  \{\alpha, \mu_{A_1}\}_D\! \approx\! \tfrac12 \sinh\psi\coth\tfrac{\mu_{B_1}}{2},\\[+.3em]
  \{\alpha, s_{B_1}\}_D \!\approx\! -\tfrac \alpha 2 \cosh\psi\coth\tfrac{\mu_{A_1}}{2}\! +\! \tfrac {s_{A_1}} 4 \sinh\psi\,(\sin\tfrac{\mu_{A_1}}{2})^{-2}, & \{\alpha, \mu_{B_1}\}_D \!\approx\! -\tfrac12 \sinh\psi\coth\tfrac{\mu_{A_1}}{2}.
  \end{array}\nonumber
\end{align}

\smallskip
\item The Dirac brackets of the variables $\mu_{A_1}, s_{A_1}, \mu_{B_1}, s_{B_1}, \alpha,\psi$ with the momentum and angular momentum variables $p^a_X,j^a_X$  ($X\neq\{A_1,B_1\}$)  of the remaining holonomies are given by
\begin{align}\label{slikeas}
  &\{\psi, p_X^a\}_D \approx 0, &
  &\{\psi, j_X^a\}_D \approx \tidadi{X}{^a_b} q_\infty^b, \\
  &\{\alpha, p_X^a\}_D \approx -\tensor{\ee}{^a_b_c} q_\infty^b p_X^c, &
  &\{\alpha, j_X^a\}_D \approx -\tensor{\ee}{^a_b_c} q_\infty^b j_X^c + \tidadi{X}{^a_b} r_\infty^b, \nonumber\\
  &\{\mu_{A_1}, p_X^a\}_D \approx 0, &
  &\{\mu_{A_1}, j_X^a\}_D \approx \tidadi{X}{^a_b} q_{A_1}^b, \nonumber\\
  &\{s_{A_1}, p_X^a\}_D \approx -\tensor{\ee}{^a_b_c} q_{A_1}^b p_X^c, &
  &\{s_{A_1}, j_X^a\}_D \approx -\tensor{\ee}{^a_b_c} q_{A_1}^b j_X^c, \nonumber\\
  &\{\mu_{B_1}, p_X^a\}_D \approx 0, &
  &\{\mu_{B_1}, j_X^a\}_D \approx \tidadi{X}{^a_b} q_{B_1}^b, \nonumber\\
  &\{s_{B_1}, p_X^a\}_D \approx -\tensor{\ee}{^a_b_c} q_{B_1}^b p_X^c, &
  &\{s_{B_1}, j_X^a\}_D \approx  - \tensor{\ee}{^a_b_c} q_{B_1}^b j_X^c+\tidadi{X}{^a_b} r_{B_1}^b,\nonumber
\end{align}
where:
\begin{align}\label{helpvecssl}
  &\bq_{A_1} \equiv \frac{\be_1}{2}, \;\;
  \bq_{B_1} \equiv \frac12\begin{pmatrix}
    \sinh\psi \\
    \cosh\psi \\
    \sinh\psi\coth\tfrac{\mu_{A_1}}{2}
  \end{pmatrix}, \;\;\bq_\infty \equiv -\frac14 \begin{pmatrix}
    \cosh\psi \coth\tfrac{\mu_{B_1}}{2} - \coth\tfrac{\mu_{A_1}}{2} \\
    \sinh\psi \coth\tfrac{\mu_{B_1}}{2} \\
    \cosh\psi \coth\tfrac{\mu_{A_1}}{2} \coth\tfrac{\mu_{B_1}}{2} - 1
  \end{pmatrix}, \nonumber\\
    &\br_\infty \equiv \frac{s_{A_1}\cosh\psi}{8\sinh^2\tfrac{\mu_{A_1}}{2}} \begin{pmatrix}
      -1/\cosh\psi \\
      0 \\
     \coth\tfrac{\mu_{B_1}}{2}
    \end{pmatrix} +
    \frac{s_{B_1}\cosh\psi}{8\sinh^2\tfrac{\mu_{B_1}}{2}} \begin{pmatrix}
      1 \\
      \tanh\psi \\
      \coth\tfrac{\mu_{A_1}}{2}
    \end{pmatrix} -
    \frac{\alpha\sinh\psi}{4\tanh\tfrac{\mu_{B_1}}{2}} \begin{pmatrix}
      1\\
      \coth\psi \\
      1
    \end{pmatrix},
\nonumber \\
  &\br_{B_1} \equiv \frac{\alpha}{2}\begin{pmatrix}
    \cosh\psi \\
    \sinh\psi \\
    \cosh\psi\coth\tfrac{\mu_{A_1}}{2}
  \end{pmatrix} -
  \frac{s_{A_1}\sinh\psi\,\be_2}{4\sinh^2\tfrac{\mu_{A_1}}{2}}.
\end{align}

\smallskip
\item The Poisson brackets of the momentum and angular momentum variables $p^a_X,j^a_X$ of the non-gauge-fixed holonomies  are given by  \eqref{ppdirac}, \eqref{pjdirac}, \eqref{jjdirac} and \eqref{V-definition} with
\begin{align}\label{handle-spacelike-wmdef}
  %\begin{aligned}
    \bw&\!=\!-\frac{1}{2\sinh\psi}\begin{pmatrix}
      \coth\tfrac{\mu_{A_1}}{2}\cosh\psi\! -\! \coth\tfrac{\mu_{B_1}}{2} \\
      2\coth\tfrac{\mu_{A_1}}{2}\sinh\psi \\
      \cosh\psi \!-\! \coth\tfrac{\mu_{A_1}}{2}\coth\tfrac{\mu_{B_1}}{2}
    \end{pmatrix}, \;\; \bm\!=\!\frac \alpha 2 \frac{\partial \bw}{\partial\psi}\!+\!\frac {s_A} 2 \frac{\partial \bw}{\partial\mu_A}\!+\!\frac {s_B} 2 \frac{\partial \bw}{\partial\mu_B}. %\\
    %\bm&=
    %  \frac{s_{A_1}}{8\sinh\psi\sinh^2\tfrac{\mu_{A_1}}{2}} \begin{pmatrix}
    %      \cosh\psi \\
    %      2\sinh\psi \\
    %      -\coth\tfrac{\mu_{B_1}}{2}
    %    \end{pmatrix} -
     % \frac{s_{B_1}}{8\sinh\psi\sinh^2\tfrac{\mu_{B_1}}{2}} \begin{pmatrix}
     %     1 \\
      %    0 \\
      %    \coth\tfrac{\mu_{A_1}}{2}
      %  \end{pmatrix} +
      %\frac{\alpha}{2} \frac{\partial \bw}{\partial\psi}.\nonumber
  \end{align}
%\end{equation}
\end{enumerate}
\end{theorem}
The Dirac brackets obtained from the two handle gauge fixing conditions exhibit many structural similarities with the one  for the particle gauge fixing condition in Theorem \ref{thm:particledirac}.
Firstly, the Dirac brackets of the momenta and angular momenta of the non-gauge-fixed holonomies take an analogous form and are again given in terms of a matrix $V$ as in  \eqref{V-definition} and a vector $\bm$ defined by \eqref{handle-timelike-wmdef}, \eqref{handle-spacelike-wmdef}. As in the particle case, the matrix $V$ depends only on the parameters $\mu_{A_1}, \mu_{B_1}$ and $\psi$ and is therefore given as a function of the Lorentzian components of the non-gauge-fixed holonomies.  However, in contrast to the particle gauge fixing condition, the vector $\bm$ is now given as a linear combination of angular momenta, and does not have a component which depends only on the Lorentzian part of the holonomies. This is due to the fact that the spin variables $s_{A_1}, s_{B_1}$ in \eqref{handle-timelike-wmdef}, \eqref{handle-spacelike-wmdef}  are dynamical variables, not external parameters as the spin variables $s_1,s_2$ of the particles.

This implies that we can identify the momentum variables of the non-gauge-fixed holonomies with functions on
 the group $PSL(2,\RR)^{n+2g-2}$ and their angular momenta  with vector fields on the same group. The structure of the Dirac bracket is thus similar to the original bracket in Theorem \ref{thm:frtheorem}: The Dirac bracket of two momenta vanishes. The Dirac bracket of momenta with angular momenta is given by the action of vector fields on functions. And the Dirac bracket of two angular momenta corresponds to the Lie bracket of the associated vector fields. In this description, the parameters $\mu_{A_1},\mu_{B_1},\psi$ are given as functions of the Lorentzian components of the non-gauge-fixed holonomies.  The parameters $s_{A_1}, s_{B_1},\alpha$  depend on both, their Lorentzian and translational components.

As in the particle gauge fixing, the parameters $\alpha,\psi$ Poisson-commute with each other.
Equations \eqref{tlikeas}, \eqref{slikeas} imply that the  variable $\psi$ generates global translations of the non-gauge-fixed holonomies in the direction of the vector $\bq_\infty$. However, in contrast to the particle case, where this  translation was in the direction of the total momentum $\hat\bp_R$, one can show that the vector $\bq_\infty$ in \eqref{helpvecstl}, \eqref{helpvecssl} is not parallel to the total momentum $\bp_{K_1}$ associated with the gauge-fixed handle via
$
u_{K_1}=[u_{B_1}, u_{A_1}^\inv]=\exp(-p_{K_1}^a J_a).
$

Similarly, equations \eqref{tlikeas}, \eqref{slikeas} imply that the  variable $\alpha$ generates global rotations around $\bq_\infty$ and global translations in the direction of the vector $\br_\infty$ defined in   \eqref{helpvecstl}, \eqref{helpvecssl}. Neither of these vectors coincides with the total momentum vector $\bp_{K_1}$ of the handle.

The main difference in relation to the particle gauge fixing condition is that the parameters $\mu_{A_1}, \mu_{B_1}, s_{A_1}, s_{B_1}$ are dynamical and do not Poisson-commute with the other phase space variables. Instead, it follows from \eqref{tlikeas}, \eqref{slikeas} that the variable $\mu_{A_1}$ generates global translations of all non-gauge-fixed holonomies in the direction of $\hat \bp_{A_1}=e_1$. The variable $s_{A_1}$ generates rotations around  $\hat \bp_{A_1}=e_1$. The variable $\mu_{B_1}$ generates global translations in the direction of  the vector $\bq_{B_1}$ in   \eqref{helpvecstl}, \eqref{helpvecssl}. However, this vector does not coincide with the vector $\hat \bp_{B_1}$. Instead, it is given as the sum
$\bq_{B_1}=\hat\bp_{B_1}+\coth\tfrac{\mu_{A_1}} 2 \hat\bp_{A_1}\wedge\hat\bp_{B_1}.
$
Similarly, the spin variable $s_{B_1}$ generates a combination of global rotations around $\bq_{B_1}$ and global translations in the direction of the vector $\br_{B_1}$ in \eqref{helpvecstl}, \eqref{helpvecssl}.

%%%%%%%%%%%%%%%%%%%%%%%%%%%%%%%%%%%

\section{Interpretation}
\label{sec:interpretation}

\subsection{Gauge fixing with particles}
\label{sec:particleinterpretation}

Having derived the Dirac bracket of the gauge-fixed system, we now
investigate its interpretation in terms of spacetime geometry.
We start by considering the Dirac bracket in Theorem \ref{thm:particledirac} that results from the particle gauge fixing condition \eqref{partconstraints}.
For this we
recall that the constraint \eqref{jpconst}, which is a Casimir function with
respect to the Dirac bracket, relates the variables associated with the two
gauge-fixed particles to the variables of the remaining particles and handles:
\begin{align}\label{const2}
(\exp(p_R^cJ_c), \bj_R)=M_2M_1\approx M_3^\inv\cdots M_n^\inv[A_1^\inv,B_1]\cdots [A_g^\inv,B_g].
\end{align}
The three-vector $\bp_R$ can thus be expressed as a function of the Lorentzian
components of the residual holonomies. The three-vector $\bj_R$ can be
identified with a vector field on $PSL(2,\RR)^{n-2+2g}$ and is given in terms of the
angular momentum vectors of the residual variables by
\begin{equation}\label{jrdef}
  \bj_R\approx
    \sum_{i=3}^n \Ad(u_{M_3}^\inv \dots u_{M_{i-1}}^\inv) \bj_\mi +
    \sum_{i=1}^g \Ad(u_{M_3}^\inv \cdots u_{M_n}^\inv u_{K_1}^\inv\cdots u_\ki^\inv) \bj_\ki,
\end{equation}
with $u_{K_i}$ and $\bj_{K_i}$ as in \eqref{handlej}.
As the first two particles were used to determine the reference frame of the
observer and their holonomies are fixed, we can interpret the non-gauge-fixed
particles and handles as a dynamical system embedded in a background geometry
that is specified by the two gauge-fixed particles.  The variables $\bp_R$ and
$\bj_R$ defined in \eqref{mupres} and \eqref{jresconcrete} can be viewed as the
total momentum and angular momentum of this residual system.  The phase space
transformations generated by these variables correspond to the effective
symmetries of the residual system.

\subsubsection{Conical symmetry via gauge fixing with particles}
\label{sec:particlecone}

To develop a precise understanding of these effective symmetries and of the
impact of gauge fixing, we compare the global  symmetries of a
non-gauge-fixed system with $n-2$ particles and the original Poisson bracket from Theorem \ref{thm:frtheorem} to the
effective symmetries of the Dirac bracket.
We consider the Fock-Rosly bracket for $n-2$ particles labelled by
holonomies $M_3,\dots,M_n$ and $g$ handles with associated holonomies
$A_1,B_1,\dots,A_g,B_g$ with the ordering of Theorem \ref{thm:frtheorem}. This
corresponds to erasing the loops $m_1,m_2$
from the set of generators of $\pi_1(S_{g,n})$ in Figure \ref{fig:fundamental-group}.  A short calculation shows that  the associated  variables $\bp_R$ and $\bj_R$ defined by \eqref{const2}, \eqref{jrdef} take the role of the variables
$\bp_C$ and $\bj_C$ for the Fock-Rosly bracket on $P_3^{n+2g}$.  Their Poisson
brackets thus take the form \eqref{constalg}:
\begin{equation}
  \{p_R^a,p_R^b\}=0, \qquad
  \{j^a_R,p^b_R\}=-\tensor{\ee}{^a^b_c}p^c_R, \qquad
  \{j^a_R,j^b_R\}=-\tensor{\ee}{^a^b_c}j^c_R.
\end{equation}
Their brackets with the momenta and angular momenta of the residual
particles and handles  are given by  \eqref{gtrafos}.  So for any $\bv\in\RR^3$
that does not depend on the phase space variables, we have for all $X \in \{M_3,
\dots, B_g\}$:
\begin{equation}\label{asfr}
\begin{aligned}
  \{\bv\cdot\bjres, \bp_X\}&=\bv\wedge\bp_X, &\qquad
  \{\bv\cdot\bpres, \bp_X\}&=0, \\
  \{\bv\cdot\bjres, \bj_X\}&=\bv\wedge\bj_X, &\qquad
  \{\bv\cdot\bpres, \bj_X\}&=\idadi{X}T^\inv(\bpres)\bv,
\end{aligned}
\end{equation}
The effective symmetries of the non-gauge-fixed system with the Fock-Rosly bracket  therefore correspond to the action of the Poincar\'e group $P_3$ on three-dimensional Minkowski space.
Hence, we can interpret this system as a (2+1)-spacetime that
is effectively Minkowskian.

To compare this with the effective symmetries of the gauge-fixed system, we
determine the Dirac bracket of its total momentum $\bp_R$ and angular momentum
$\bj_R$ with the momenta and angular momenta of the non-gauge-fixed particles
and handles.  The form of these expressions on the constraint surface follows directly from
 equations \eqref{p2brack} and from expressions
\eqref{mupres}--\eqref{jsxres} for $\bp_R$ and $\bj_R$ in terms of the
parameters $\psi$ and $\alpha$. We obtain the following corollary.

\begin{corollary}\label{lemma:assym}
The components of the total momentum and angular momentum of the residual system
Poisson-commute weakly:
\begin{equation}\label{RR-dirac-brackets}
  \{p^a_{R}, p^b_{R}\}_D \approx \{j^a_{R}, p^b_{R}\}_D \approx \{j^a_{R}, j^b_{R}\}_D \approx 0.
\end{equation}
For all $X\in\{M_3,\dots,M_n,A_1,\dots,B_g\}$ and $\bv\in\RR^3$, we have:
\begin{subequations}\label{RX-dirac-brackets}
  \begin{align}
    \label{pRpX-dirac-bracket}
    \{\bv\cdot\bp_R, \bp_X\}_D &\approx 0, \\
    \label{pRjX-dirac-bracket}
    \{\bv\cdot\bp_R, \bj_X\}_D &\approx \tau(\bv)\,\idadi{X}\hbp_R, \\
    \label{jRpX-dirac-bracket}
    \{\bv\cdot\bj_R, \bp_X \}_D &\approx
      \phi(\bv)\,\hat\bp_R\wedge\bp_X, \\
    \label{jRjX-dirac-bracket}
    \{\bv\cdot\bj_R, \bj_X\}_D &\approx
      \phi(\bv)\,\hat \bp_R\wedge\bj_X -
      \idadi{X}[\sigma(\bv)\hat \bp_{R}+\phi(\bv) \hat\bp_{R}\wedge \bx_{R}],
  \end{align}
\end{subequations}
where $\phi(\bv)$, $\tau(\bv)$, $\sigma(\bv)$ are linear functions of $\bv$
given by expression \eqref{p2brack} for $\xi_R$
 and\begin{equation}\label{asfuncpart}
  \begin{gathered}
    \tau(\bv)\!=\!\frac{\bv\cdot (\partial \bp_{R}/\partial\psi)}{\hat\bp_{R}\cdot (\partial \bp_{R}/\partial\psi)}, \quad
    \phi(\bv)\!=\!\frac{\bv\cdot (\partial \bj_{R}/\partial\alpha)}{\hat\bp_{R}\cdot (\partial \bj_{R}/\partial\alpha)}, \quad
    \sigma(\bv)\!=\!\frac{\bv\cdot(\partial\bj_{R}/\partial\psi)}{\hat\bp_{R}\cdot(\partial\bj_{R}/\partial \alpha)} -\xi_R\,\phi(\bv).
  \end{gathered}
\end{equation}
\end{corollary}

Equations \eqref{pRpX-dirac-bracket} and \eqref{pRjX-dirac-bracket} state that
the phase space transformations generated by the components of the total
momentum $\bp_R$ are global translations in the direction of $\bp_R$.
Equations \eqref{jRpX-dirac-bracket} and \eqref{jRjX-dirac-bracket} imply that
the phase space transformations generated by the components of the total angular
momentum $\bj_R$ are global translations in the direction of $\bp_R$ together
with rotations around the geodesic $g_R(t)=t\hbp_R+\bx_R$, $\bx_R\cdot\bp_R=0$.  Their action on points  in Minkowski space is given by
\begin{equation}
  \by\mapsto \bx_R+ \Ad(u_R(\phi))(\by-\bx_R)+\sigma\hat\bp_R, \qquad
  u_R(\phi)=\exp(\phi\,\hat p_R^aJ_a).
\end{equation}
The term
$\phi(\bv)\bp_{R}\wedge \bx_{R}$ in \eqref{jRjX-dirac-bracket} arises from the fact that this
geodesic does not go through the origin but has an offset $\bx_{R}$.
The parameter $\phi(\bv)$ defines the angular
velocity of the rotation generated by the angular momentum vector $\bj_R$.  The
parameters $\tau(\bv)$ and $\sigma(\bv)$ define the relative velocity of the
translations generated by, respectively, the momentum $\bp_R$ and the angular
momentum $\bj_R$.

It follows from the
discussion in Sections \ref{sec:3dgrav} and \ref{sec:partgfix}  that
these translations and rotations are precisely the symmetries of the cone whose axis is given by the
geodesic $g_R$. This is the cone associated to the two gauge-fixed particles via
\eqref{mupres}--\eqref{jsxres}.
The effective symmetry group of the Dirac bracket is therefore the
two-dimensional abelian symmetry group of  a cone in Minkowski space, which is generated
by a rotation around its axis and a translation in the direction of its axis.
This allows us to
interpret the system as a (2+1)-spacetime which is effectively conical.  As detailed in Section \ref{sec:3dgrav}, an element of the Poincar\'e group defines a cone in Minkowski space. Corollary \ref{lemma:assym} states that  this cone is the one given by the product  $M_2M_1$ of the holonomies of the gauge fixed particles. 
 Its deficit angle and time shift are dynamical variables given by
the relative velocity $\tanh \psi$ and the minimal distance $\alpha$ of the
gauge-fixed particles or, equivalently, the total mass $\mu_R$ and spin $s_R$ of
the non-gauge-fixed system.

The particle gauge fixing procedure therefore has a clear interpretation in terms of
spacetime geometry: It describes the transition from an effectively
Minkowskian (2+1)-spacetime associated with the original bracket to an
effectively conical spacetime associated with the Dirac bracket.  The
geometry of the cone is determined by the mass and spin parameters of the two
gauge-fixed particles and by the two dynamical variables that characterise their
relative motion, their relative velocity $\tanh \psi$ and their minimum distance
$\alpha$.  The former defines the opening angle of the cone and the latter its
time shift.

\subsubsection{Particle gauge fixing as a global translation}
\label{sec:transl}

To analyse the geometrical interpretation of the particle gauge fixing procedure and the
resulting Dirac bracket further, we investigate its relation to the original
bracket in Theorem \ref{thm:frtheorem}. In this context, it is natural to ask if the Dirac bracket
\eqref{dirac-non-gauge-fixed} for the variables of the non-gauge-fixed particles
and handles can be related to the original bracket \eqref{frbracket} via
 a certain coordinate transformation.  More precisely, we ask if there
exists a diffeomorphism $\Gamma: P_3^{n+2g-2}\rightarrow P_3^{n+2g-2}$,
 such that the bracket \eqref{frbracket} of
the transformed momentum and angular momentum variables agrees with their Dirac
bracket on the constraint surface:
\begin{align}\label{goal}
  \{f\circ \Gamma, g\circ \Gamma\} \approx \{f,g\}_{D}\circ \Gamma \qquad
    \forall f,g\in\cif(P_3^{n+2g-2}).
\end{align}
Note that this condition does {\em not} imply that Fock and Rosly's Poisson
structure is Poisson-equivalent to the gauge-fixed Poisson structure in Theorem
\ref{thm:particledirac}. This is because Fock and Rosly's Poisson
bracket does {\em not} restrict to a Poisson structure on the constraint
surface. Moreover, since we inverted the Dirac matrix only on the constraint
surface, it cannot be expected that identity \eqref{goal} holds globally.

To determine if there exists a transformation $\Gamma:P_3^{n+2g-2}\rightarrow
P_3^{n+2g-2}$ that satisfies condition \eqref{goal}, we note that formula
\eqref{pjdirac} is consistent with a transformation
that affects only the angular
momentum variables $\bj_X$, $X\in\{M_3,\dots,M_n\}$, and transforms them by a
global translation that depends on the vector $\bj_R$, on the total mass $\mu_R$
and on external parameters that Poisson-commute with all functions in
$\cif(P_3^{n+2g-2})$. We obtain the
following theorem.

\begin{theorem}[Particle gauge fixing as a global translation]\label{thm:translation}\mbox{}

Consider the Poisson manifold $M_{FR}=(P_3^{n+2g-2},\{\,,\,\})$,
%and
%$M_D=(P_3^{n+2g-2}, \allowlinebreakhere \{\,\,\}_D)$
where $\{\,,\,\}$ is the
Poisson bracket \eqref{frbracket}
%and $\{\,\,\}_D$ the Dirac bracket
%\eqref{dirac-non-gauge-fixed},
%both
restricted to the system with the first two
particles removed.
Denote by $\bp_R,\bj_R$ the total momentum and angular momentum defined via
\begin{equation}
(\exp(p_R^a), \bj_R)=M_3^\inv\cdots M_n^\inv[A_1^\inv,B_1]\cdots[A_g^\inv,B_g].
\end{equation}
Introduce external parameters $\mu_1,\mu_2, s_1,s_2$ which Poisson-commute with all functions in $\cif(P_3^{n+2g})$ and define  $\psi\in\cif(P_3^{n+2g-2})$ as a function of the total mass $\mu_R$ and of the parameters $\mu_1,\mu_2$ through equations \eqref{mupres}. Define  $\alpha\in\cif(P_3^{n+2g-2})$ as a function of $\mu_1,\mu_2,s_1,s_2$ and the total spin $s_R$ through equation \eqref{srdef}.

Let $\Gamma: P_3^{n+2g-2}\rightarrow P_3^{n+2g-2}$ be a global  translation:
\begin{equation}\label{sigdef}
  \Gamma: \begin{cases}
    u_X \mapsto u_X, \\
    \bj_X \mapsto \bj_X+\idadi{X}\bt,
  \end{cases}\qquad\forall X\in\{M_3,\dots,M_n,A_1,\dots,B_g\}.
\end{equation}
Suppose the translation vector $\bt$ is of the form $\bt=-V\bj_R+\ba$, where
 the $3\times 3$-matrix $V$ is given as a function of the variables $\mu_R, \mu_1,\mu_2$ by \eqref{V-definition}, \eqref{wmdef} and
the three-vector $\ba$ is a function of $\mu_1, \mu_2, \mu_R, s_1, s_2, s_R$ which is given below in the proof.
Then the map $\Gamma:P_3^{n+2g-2}\rightarrow P_3^{n+2g-2}$  satisfies
\begin{equation}
  \{f\circ \Gamma, g\circ \Gamma\}\approx\{f,g\}_{D}\circ \Gamma\quad\forall f,g\in\cif(P_3^{n+2g-2}),
\end{equation}
where $\{\,,\,\}_D$ is the Dirac bracket for the particle gauge fixing derived in Theorem \ref{thm:particledirac} and $\approx$ denotes equality on the constraint surface $C'\subset P_3^{n+2g-2}$ defined by
conditions \eqref{mupres}.
\end{theorem}

\begin{proof}
The proof is a direct but lengthy calculation. We use the coordinates from
Theorem \ref{thm:frtheorem} and start by determining the Poisson brackets of
momentum and angular momentum variables. From the definition of the map
$\Gamma$, we have
\begin{equation}
\{j^a_X\circ \Gamma, p^b_Y\circ \Gamma\}=\{j^a_X, p^b_Y\}+\tidadi{X}{^a_c}\{t^c, p^b_Y\}
\end{equation}
for $X,Y\in\{M_3,\dots,M_n,A_1,\dots,B_g\}$. With the identities
\begin{align}\label{helpids1}
&\{\mu_R, p^b_Y\}=0,\qquad \{t^c, p^b_Y\}=-\tensor{V}{^c_d}\{j^d_R, p^b_Y\}=\tensor{V}{^c_d}\tensor{\ee}{^d^b_f}\,p^f_Y,
\end{align}
which follow directly from the expressions for Fock and Rosly's Poisson
structure in Theorem \ref{thm:frtheorem}, we obtain agreement with formula
\eqref{pjdirac} in Theorem \ref{thm:particledirac}.

To determine the brackets of the angular momentum variables, we use the identity
\begin{align}\label{helpident}
\{j^a_X,j^b_Y\}\circ \Gamma
&=\{j^a_X,j^b_Y\}-\{j^a_X, \tensor{\Ad(u_Y^\inv)}{^b_d}\}t^d
+\{j^b_Y, \tensor{\Ad(u_X^\inv)}{^a_c}\}t^c\nonumber\\
&-\tidadi{X}{^a_c}\tidadi{Y}{^b_d}\,\ee^{cdf}\, t_f,
\end{align}
and obtain
\begin{align}\label{helpbr}
\{j^a_X\circ \Gamma, j^b_Y\circ\Gamma\}&=\{j^a_X,j^b_Y\}\circ \Gamma+\tidadi{Y}{^b_d}\{j^a_X, t^d\}-\tidadi{X}{^a_c}\{j^b_Y, t^c\}\nonumber\\
&-\tidadi{X}{^a_c}\{t^c,\tensor{\Ad(u_Y^\inv)}{^b_d}\}t^d+\tidadi{Y}{^b_d}\{t^d,\tensor{\Ad(u_X^\inv)}{^a_c}\}t^c\nonumber\\
&+\tidadi{X}{^a_c}\tidadi{Y}{^b_d}(\{t^c,t^d\}+\ee_{cdf}t^f).
\end{align}
%To relate this expression to the Dirac bracket
%we use the auxiliary identities
%\begin{align}
%&\{t^c,t^d\}=-V^{cg}V^{dk}\ee_{gkf}j^f_R\qquad
%\{j^a_X, V^{bc}\}=\frac{\partial V^{bc}}{\partial \mu_R}\{j^a_X,\mu_R\}\qquad \{j^a_X, a^b\}=\frac{\partial a^b}{\partial\mu_R}\{j^a_X,\mu_R\}\nonumber\\
%&\{t^d,\Ad(u_X^\inv)^a_{\;\;c}\}=V^{dg}(\ee_{g}^{\;\;ak}\Ad(u_X^\inv)_{kc}+\ee_{gck}\Ad(u_X^\inv)^{ak})\qquad
%\{j^a_X,\mu_R\}=-(1-\Ad(u_X^\inv))\hat\bp_R,\nonumber
%\end{align}
%which follow from expressions \eqref{singlebr}, \eqref{mixbr}, \eqref{handlebr} for the Fock-Rosly bracket and the definition of the map $\Sigma$. After some further computations, we obtain
After some further computations this reduces to
\begin{multline}\label{tildeb}
\{j^a_X\circ\Gamma, j^b_Y\circ \Gamma\}=\{j^a_X,j^b_Y\}\circ \Gamma+\tidadi{X}{^a_c}\tensor{V}{^c_d} \,\tensor{\ee}{^d^b_k} (j^k_Y\circ\Gamma)\\
-\tidadi{Y}{^b_c}\tensor{V}{^c_g}\tensor{\ee}{^g^a_k}(j^k_X\circ\Gamma)+\idadi{X}^{ac}\idadi{Y}^{bd}\ee_{cdf}\tilde m^f,
\end{multline}
where the three-vector $\tilde \bm$ takes the form
\begin{equation}
\tilde \bm=\left(\tfrac 1 4 (\bw^2+1)+\tfrac 1 2 \hat\bp_R\cdot\frac{\partial\bw}{\partial\psi}\right)\bj_R- \tfrac {s_R} 2 \frac{\partial\bw}{\partial \mu_R}
+\tfrac 1 2 \bw\wedge\ba-\hat\bp_R\wedge\frac{\partial \ba}{\partial\mu_R}.
\end{equation}
On the constraint surface $C'$, the three-vector $\hat\bp_R$ is given by \eqref{pres}, which implies that the factor in front of $\bj_R$ vanishes.  If we set
\begin{align}\label{adef}
\ba=\frac{\sin\tfrac{\mu_1} 2\sin\tfrac{\mu_2} 2\sinh\psi}{2\sin\tfrac{\mu_R} 2}\bigg[&\frac{ s_1\,\hat\bp_{M_2}\wedge\hat\bp_R+s_2\,\hat\bp_{M_1}\wedge\hat\bp_{R}}{\sin^2\tfrac{\mu_1} 2}\\
+&\bigg(s_1(\cot\tfrac{\mu_1} 2\!+\!\cot\tfrac{\mu_2} 2 \cosh\psi)\!+\!s_2(\cot\tfrac{\mu_2} 2\!+\!\cot\tfrac{\mu_1} 2\cosh\psi)\bigg)\;\hat\bp_R\wedge\frac{\partial\bw}{\partial\psi}\bigg],\nonumber
\end{align}
with $\hat\bp_{M_1}=\be_0$, $\hat\bp_{M_2}=\cosh\psi\, \be_0+\sinh\psi\, \be_1$ and $\bw$ defined via \eqref{wmdef}, we obtain $\tilde \bm\approx\bm$ with $\bm$ given by \eqref{wmdef}. This proves the claim.
\end{proof}

Theorem \ref{thm:translation} states that the Dirac bracket of the residual
system, \ie the system without the two gauge-fixed particles, can be obtained
from its Fock-Rosly bracket by applying a global translation \eqref{sigdef} that
depends on the total mass $\mu_R$ and on the total angular momentum $\bj_R$ of
the residual system.  Under this translation the total angular momentum
transforms as
\begin{align}\label{wmatrix}
\Gamma: \bj_R\mapsto W\bj_R+\idadi{R}\ba,\qquad W=\mathds{1}-\idadi{R}V.
\end{align}
A short calculation shows that the matrix $W$ takes the form $W^{ab}= u^a\hat
p_R^b$ with a spacelike vector $\bu$ that satisfies $\bu\cdot \hat
\bp_R=1$. This implies that on the constraint surface defined by \eqref{mupres}, $W^T$ is a projector onto $\text{Span}(\hat\bp_R)$. This ensures
that the Lorentz transformations generated by $\bj_R$ involve only rotations
around the axis of the cone defined by the gauge-fixed particles. The gauge
fixing procedure can thus be understood as a global translation which modifies
the total angular momentum in such a way that the associated Lorentz transformations are reduced to a rotation around the axis of the cone.

Similarly, the specific form of the vector $\ba$ ensures that the translations
generated by the total angular momentum $\bj_R$ are no longer the full set of
translations in $\mathbb R^3$ but, on the constraint surface, reduce to
translations in the direction of the cone's axis.  Note, however, that the agreement of
the resulting bracket with the Dirac bracket defines the vector $\ba$ in Theorem
\eqref{thm:translation} only up to translations in the direction of the momentum
$\hat\bp_R$.

This provides a simple and direct interpretation of the gauge fixing procedure
and the Dirac bracket in terms of the geometry of the associated
spacetime. Gauge fixing can be understood as a global translation of the
non-gauge-fixed particles and handles which is such that the translated total
angular momentum generates precisely the symmetries of the cone defined by the
two gauge-fixed particles.

The total mass $\mu_R$ of the residual system plays the role of a total energy variable and can be viewed as the Hamiltonian of the system. The associated transformations are translations in a preferred time direction, which is given by the axis of the effective cone.
The variable $s_R$ plays the role of a total angular momentum variable of the system. The associated phase space transformations are  rotations around the axis of the cone.

Note also that Theorem \ref{thm:translation}  allows one to extend the Dirac bracket beyond the constraint surface. While the formulas in Theorem \ref{thm:particledirac} are only valid on the constraint surface defined by \eqref{mupres}, the bracket resulting from the global translation in Theorem \ref{thm:translation} is defined for all values of the variables $\bp_X,\bj_X$, $X\in\{M_3,...,M_n,A_1,...,B_g\}$.

\subsection{Gauge fixing with handles}

\label{handleinterpret}

As in the case of the particle gauge fixing, we will now give a physical interpretation of the Dirac brackets for handles in terms of the geometry of the residual variables.

As the degrees of freedom of the first handle were used to determine the reference frame of the
observer, they can be eliminated from the description and expressed in terms of the momenta and angular momenta of the residual handles via the constraint
\begin{align}\label{handexp}
(\exp(p_R^c J_c),\bj_R)=[B_1,A_1^\inv]\approx[A_2^\inv,B_2]\cdots [A_g^\inv,B_g]
\end{align}
together with
 \eqref{pkhandle}. This allows us to interpret  the three-vectors $\bp_R$, $\bj_R$ in \eqref{handexp} as a total momentum and  angular momentum vector of the residual system, \ie the handles with holonomies $A_2,B_2,...,A_g,B_g$.
 The phase space
transformations generated by these variables can be viewed as  the effective
symmetries of the residual system.

\subsubsection{Poincar\'e symmetry via gauge fixing with handles}

From Theorem \ref{thm:handledirac} it follows directly that these effective symmetries are the ones of Minkowski space. By taking the partial derivatives of $\bp_R,\bj_R$ with respect to the parameters of the gauge-fixed handle, we obtain for all $f\in \cif(P_3^{n+2g-2})$
\begin{align}
\{p^a_R, f\}_D &\approx \frac{\partial p^a_R}{\partial \psi} \{\psi, f\}_D+\frac{\partial p^a_R}{\partial \mu_{A_1}} \{\mu_{A_1}, f\}_D+\frac{\partial p^a_R}{\partial \mu_{B_1}} \{\mu_{B_1}, f\}_D,\\
\begin{split}
\{j^a_R, f\}_D &\approx \frac{\partial j^a_R}{\partial \psi} \{\psi, f\}_D+\frac{\partial j^a_R}{\partial \mu_{A_1}} \{\mu_{A_1}, f\}_D+\frac{\partial j^a_R}{\partial \mu_{B_1}} \{\mu_{B_1}, f\}_D\\
&+\frac{\partial j^a_R}{\partial \alpha} \{\alpha, f\}_D+\frac{\partial j^a_R}{\partial s_{A_1}} \{s_{A_1}, f\}_D+\frac{\partial j^a_R}{\partial s_{B_1}} \{s_{B_1}, f\}_D.
\end{split}
\end{align}
Using the expressions \eqref{tlikeas}, \eqref{slikeas} for the Dirac bracket in Theorem \ref{thm:handledirac}, we find that the components of the total momentum vector $\bp_R$ generate three linearly independent translations in the direction of the vectors $\bq_\infty$, $\bq_{A_1}$, $\bq_{A_2}$ in \eqref{helpvecstl}, \eqref{helpvecssl}.
Similarly, the components of the angular momentum vector $\bj_R$ generate three independent Lorentz transformations with axes $\bq_\infty,\bq_{A_1}, \bq_{B_1}$ and two translations in the direction of the vectors $\br_\infty,\br_{B_1}$ given by \eqref{helpvecstl}, \eqref{helpvecssl}.

The effective symmetries of the handle gauge-fixed system are therefore the ones of  Minkowski space, and its effective symmetry group is the Poincar\'e group $P_3$ in three dimensions.
This distinguishes the handle gauge fixing from the particle gauge fixing, for which the effective symmetries were conical. It is a consequence of the fact that the gauge-fixed handle has six residual dynamical degrees of freedom: the variables $\mu_{A_1}, \mu_{B_1}, s_{A_1}, s_{B_1}, \psi, \alpha$, whereas a system of two gauge-fixed particles has only two such degrees of freedom, the parameters $\psi, \alpha$ which define the deficit angle and time shift of the cone.

\subsubsection{Handle gauge fixing as a global translation}

As in the case of the particle gauge fixing, the form of the Dirac bracket in Theorem \ref{thm:handledirac} suggests that the Dirac bracket for the handle gauge fixings could be obtained from the Fock-Rosly bracket by applying a global translation.
However, in contrast to the particle case, one would expect this relationship to hold globally. For the particle gauge fixing, the constraint surface defined by \eqref{mupres}--\eqref{jsxres} is a $(6[n-2]+12g-4)$-dimensional submanifold of the $(6[n-2]+12 g)$-dimensional  manifold $P_3^{n+2g-2}$. In contrast, the handle gauge fixing constraints yield a manifold of dimension $12(g-1)$, which coincides with the dimension of $P_3^{2g-2}$.  This suggests that the Dirac bracket can be obtained from the Fock-Rosly bracket for a reduced system with $g-1$ handles through a coordinate transformation. This is the content of the following theorem.

\begin{theorem}[Handle gauge fixing as a global translation]\label{thm:handletranslation}\mbox{}

Consider the Poisson manifold $M_{FR}=(P_3^{2g-2},\{\,,\,\})$,
where $\{\,,\,\}$ is the
Fock-Rosly bracket \eqref{frbracket}
restricted to the system with the first handle removed. Define the total momentum $\bp_R$ and angular momentum $\bj_R$ of the system via
\begin{equation}
(\exp(p_R^aJ_a),\bj_R)=[A_2^\inv,B_2^\inv]\cdots[A_g^\inv, B_g].
\end{equation}
Let $\Gamma: P_3^{2g-2}\rightarrow P_3^{2g-2}$ be the
smooth map that describes the transformation of the holonomies under global conjugation with a translation by  $\bt=-V\bj_R$:
\begin{equation}\label{sigdefhand}
  \Gamma: \begin{cases}
    u_X \mapsto u_X, \\
    \bj_X \mapsto \bj_X-\idadi{X}V\bj_R,
  \end{cases}\qquad\forall X\in\{A_2,\dots,B_g\},
\end{equation}
where the matrix $V$ is a function of the total momentum $\bp_R$.
Introduce variables
$\mu_{A_1}, \mu_{B_1}, s_{A_1}, \allowlinebreakhere s_{B_1}, \psi, \alpha \in\cif(P_3^{2g-2})$ by defining them as functions of the total momentum $\bp_R=\bp_R\circ \Gamma$ and the total angular momentum $\bj_R\circ\Gamma$ through equations \eqref{pkhandle}:
\begin{align}\label{paradef}(\exp(p_R^aJ_a), \bj_R\circ\Gamma)=[B_1,A_1^\inv],\end{align}
where $B_1,A_1$ are given by \eqref{handle-timelike-gauge} for the timelike gauge fixing condition or by \eqref{handle-spacelike-gauge} for the spacelike gauge fixing condition. Suppose the matrix $V$ is of the form
$V_{ab}=\tfrac 1 2 (\eta_{ab}+\ee_{abc}w^c)$, where the three-vector $\bw$ given as a function of $\bp_R,\bj_R$ by \eqref{handle-timelike-wmdef} (timelike gauge fixing condition) or  \eqref{handle-spacelike-wmdef} (spacelike gauge fixing condition).
Then the map $\Gamma:P_3^{2g-2}\rightarrow P_3^{2g-2}$  satisfies
\begin{equation}
  \{f\circ \Gamma, g\circ \Gamma\}=\{f,g\}_{D}\circ \Gamma\quad\forall f,g\in\cif(P_3^{2g-2}),
\end{equation}
where $\{\,,\,\}_D$ is the Dirac bracket for the timelike or spacelike gauge fixing condition given in Theorem \ref{thm:handledirac}.
\end{theorem}

\begin{proof} The proof is similar to the one of Theorem \ref{thm:translation}.
For the Poisson bracket of momenta and angular momenta $j^a_X,p^b_Y$, $X,Y\in\{A_2,...,B_g\}$, we obtain
\begin{equation}
  \{j^a_X\circ\Gamma, p^b_Y\circ\Gamma\}
    %=\{j^a_X, p^b_Y\}-\tidadi{X}{^a_c}\tensor{V}{^c_g}\{j^g_R, p^b_Y\}
    =\{j^a_X, p^b_Y\}+\tidadi{X}{^a_c}\tensor{V}{^c_g}\ee^{gbf}p_f^Y,
\end{equation}
which agrees with the expression in Theorem \ref{thm:handledirac}.
To determine the brackets of the angular momentum variables, we use the identity
\begin{equation}%\label{helpident}
\begin{split}
\{j^a_X,j^b_Y\}\circ \Gamma
&=\{j^a_X,j^b_Y\}+\{j^a_X, \tensor{\Ad(u_Y^\inv)}{^b_d}\}V^{dg}j^R_g
-\{j^b_Y, \tensor{\Ad(u_X^\inv)}{^a_c}\}V^{cg}j^R_g\\
&+\tidadi{X}{^a_c}\tidadi{Y}{^b_d}\,\ee^{cdf}\, V_{fg}j^g_R,
\end{split}
\end{equation}
which yields
\begin{equation}%\label{helpbr}
\begin{split}\raisetag{1.5em}
\{j^a_X\circ \Gamma, j^b_Y\circ\Gamma\}&=\{j^a_X,j^b_Y\}\circ \Gamma+\tidadi{Y}{^b_d}\{j^a_X, t^d\}-\tidadi{X}{^a_c}\{j^b_Y, t^c\}\\
&-\tidadi{X}{^a_c}\{t^c,\tensor{\Ad(u_Y^\inv)}{^b_d}\}t^d+\tidadi{Y}{^b_d}\{t^d,\tensor{\Ad(u_X^\inv)}{^a_c}\}t^c\\
&+\tidadi{X}{^a_c}\tidadi{Y}{^b_d}(\{t^c,t^d\}+\ee_{cdf}t^f).
\end{split}
\end{equation}
%To relate this expression to the Dirac bracket
%we use the auxiliary identities
%\begin{align}
%&\{t^c,t^d\}=-V^{cg}V^{dk}\ee_{gkf}j^f_R\qquad
%\{j^a_X, V^{bc}\}=\frac{\partial V^{bc}}{\partial \mu_R}\{j^a_X,\mu_R\}\qquad \{j^a_X, a^b\}=\frac{\partial a^b}{\partial\mu_R}\{j^a_X,\mu_R\}\nonumber\\
%&\{t^d,\Ad(u_X^\inv)^a_{\;\;c}\}=V^{dg}(\ee_{g}^{\;\;ak}\Ad(u_X^\inv)_{kc}+\ee_{gck}\Ad(u_X^\inv)^{ak})\qquad
%\{j^a_X,\mu_R\}=-(1-\Ad(u_X^\inv))\hat\bp_R,\nonumber
%\end{align}
%which follow from expressions \eqref{singlebr}, \eqref{mixbr}, \eqref{handlebr} for the Fock-Rosly bracket and the definition of the map $\Sigma$. After some further computations, we obtain
After some further computations this reduces to
\begin{align}\label{tildeb2}
&\{j^a_X\circ\Gamma, j^b_Y\circ \Gamma\}=\{j^a_X,j^b_Y\}\circ \Gamma+\tidadi{X}{^a_c}\tensor{V}{^c_d} \,\tensor{\ee}{^d^b_k} (j^k_Y\circ\Gamma)\\
&\quad+\idadi{X}^{ac}\{j^b_Y\circ\Gamma, V_{cf}\}j^f_R-\idadi{Y}^{bd}\{j^a_X\circ\Gamma, V_{df}\}j^f_R\nonumber\\
&\quad-\tidadi{Y}{^b_c}\tensor{V}{^c_g}\tensor{\ee}{^g^a_k}(j^k_X\circ\Gamma)-\tfrac 1 4 (\bw^2+1)\idadi{X}^{ac}\idadi{Y}^{bd}\ee_{cdf}j_R^f.\nonumber
\end{align}
Using the formula for the Fock-Rosly bracket in Theorem \ref{thm:frtheorem}, we then derive the identities
\begin{equation}
\{j^a_X\circ\Gamma, V_{df}\}=\frac{\partial V_{df}}{\partial p_k^R}\{j^a_X\circ\Gamma, p_R^k\}=-\frac{\partial V_{df}}{\partial p^R_k}\idadi{X}^{ac}(T^\inv(\bp_R)_{ck} - V_{cg}\tensor{\ee}{_g^k^l}p^R_l),
\end{equation}
where $T^\inv(\bp_R):\RR^3\rightarrow \RR^3$ is the map defined given by \eqref{tdef}.
Inserting these expressions into \eqref{tildeb2} together with the definition of the map $T^\inv(\bp)$ and the definition of the variables $\mu_{A_1}, \mu_{B_1}, \psi$, we obtain after a lengthy computation:
\begin{align}
&\{j^a_X\circ\Gamma, j^b_Y\circ \Gamma\}=\{j^a_X,j^b_Y\}\circ \Gamma+\tidadi{X}{^a_c}\tensor{V}{^c_d} \,\tensor{\ee}{^d^b_k} (j^k_Y\circ\Gamma)\\
&\qquad +\idadi{X}^{ac}\{j^b_Y\circ\Gamma, V_{cf}\}j^f_R+\idadi{X}^{ac}\idadi{Y}^{bd}\ee_{cdf}\tilde m^f\nonumber,
\end{align}
where the three-vector $\tilde \bm$ takes the form
\begin{align}\label{mprelim}
\tilde \bm=-\tfrac 1 2 {(\bj_R\bq_{A_1})}\frac{\partial\bw}{\partial \mu_{A_1}}-\tfrac 1 2 {(\bj_R\bq_{B_1})}\frac{\partial\bw}{\partial \mu_{B_1}}-\tfrac 1 2 {(\bj_R\bq_{\infty})}\frac{\partial\bw}{\partial \psi},
\end{align}
and $\bq_{A_1}$, $\bq_{B_1}$, $\bq_\infty$ are given by \eqref{helpvecstl} (timelike gauge fixing condition) or \eqref{helpvecssl} (spacelike gauge fixing condition).

To evaluate this expression further, we make use of condition \eqref{paradef} which relates the angular momentum $\bj_R$ to the variables $\mu_{A_1}, \mu_{B_1},\psi, s_{A_1}, s_{B_1}, \alpha$. We obtain:
\begin{equation}
\bj_R\circ \Gamma=W\bj_R=-\left[\Ad(u_{B_1})-\Ad(u_{B_1}u_{A_1}^\inv)\right]\bj_{B_1}+\left[\Ad(u_{B_1})-\Ad([u_{B_1}, u_{A_1}^\inv])\right]\bj_{A_1},
\end{equation}
where $u_{B_1}$, $u_{A_1}$, $\bj_{A_1}$, $\bj_{B_1}$ are given by \eqref{handle-timelike-gauge} (timelike gauge fixing condition) or \eqref{handle-spacelike-gauge} (spacelike gauge fixing condition), and the invertible matrix $W$ is defined as in \eqref{wmatrix}.
After inverting the matrix $W$ and performing some further evaluations, we obtain
\begin{equation}
\bq_{A_1}\bj_R=-s_{A_1},\qquad \bq_{B_1}\bj_R=-s_{B_1},\qquad \bq_\infty\bj_R=-\alpha.
\end{equation}
Inserting these expressions into \eqref{mprelim} then yields the expression for the vector $\bm$ in Theorem \ref{thm:handledirac} and
proves the claim.
\end{proof}

Theorem \ref{thm:handletranslation} states that the Dirac bracket associated with the two gauge fixing conditions for handles is obtained from the Fock-Rosly bracket on $P_3^{2g-2}$ via the invertible coordinate transformation \eqref{sigdefhand}. It is therefore Poisson-equivalent to the restriction of the Fock-Rosly bracket to $g-1$ handles. However, despite this equivalence, passing from the bracket in Theorem \ref{thm:frtheorem} to the Dirac bracket in Theorem \ref{thm:handledirac} has non-trivial physical implications. Unlike the variables in the Fock-Rosly bracket, the variables $\bp_X,\bj_X$ in the Dirac bracket have a physical interpretation in terms of the geometry of the non-gauge-fixed handles. Passing from the Dirac bracket to the Fock-Rosly bracket for $g-1$ handles therefore leads to a set of variables that no longer have a direct physical interpretation.

\section{Outlook and Conclusions}
\label{sec:outlook}

In this article, we applied Dirac's gauge fixing procedure to the phase space of
(2+1)-dimensional gravity with vanishing cosmological constant. We considered spacetimes of general genus and with a general number of punctures representing massive point particles with spin.  We showed that in this context imposing gauge
fixing conditions amounts to specifying an observer in the spacetime. This leads to two types
of natural gauge fixing conditions. The first characterises
 an observer with respect to the worldlines of two point particles, the second with respect to the geometry of a handle.

We derived
explicit expressions for the associated
Dirac brackets and showed that the Dirac bracket has a  direct  interpretation in terms of spacetime geometry.
For the gauge fixing condition based on two point particles, the symmetries of the
resulting Dirac bracket are those of a cone whose opening angle and time shift
 are given, respectively,  by the relative velocity and the minimal distance of the
two  particles. The gauge fixing condition based on a handle leads to an effectively Minkowskian spacetime.

In both cases,  there is a direct relation between the Dirac bracket and Fock and Rosly's  Poisson structure for a reduced system with, respectively,  two particles or a handle removed.
The Dirac bracket is obtained from the bracket for the reduced system via a global translation. For the gauge fixing condition based on particles, this allows one to extend the Dirac bracket beyond the constraint surface. For the gauge fixing condition based on the geometry of a handle, it demonstrates that the Dirac bracket is Poisson-isomorphic
to the Fock-Rosly bracket for the reduced system.

We expect  our gauge fixing procedure to generalise straightforwardly to Euclidean (2+1)-gravity with vanishing cosmological constant. In this case,  the  Poincar\'e group in three dimensions is replaced by the Euclidean group and Fock and Rosly's Poisson structure is of a very similar form. Using the explicit description of the Poisson structure in \cite{Meusburger:2007ad}, it should also be possible in principle to generalise 
it to Lorentzian (2+1)-gravity with non-vanishing cosmological constant and to Euclidean (2+1)-gravity with negative cosmological constant. However, as  the relevant isometry groups are no longer semidirect product groups in these cases, inverting the Dirac matrix will become much more involved.  While it reduces to inverting  $3\times 3$-matrices for the case of vanishing cosmological constant, this will no longer be the case when a non-trivial cosmological constant is present. 

Given that the gauge fixing procedure involves a system with six first-class constraints and six gauge fixing conditions, the resulting Dirac bracket is surprisingly simple. This hints at the presence of underlying mathematical structures that give rise to these simplifications.
The results of \cite{Buffenoir:2005zi}, which investigates  the gauge-invariant phase space in $SL(2,\CC)$-Chern-Simons theory with punctures, indicate that these could be  dynamical quantum groups and the associated classical structures. We intend to explore the relation between the Dirac bracket and dynamical quantum group symmetries in a future work.
Generally, it would be interesting to compare our gauge fixing to the results in \cite{Buffenoir:2005zi}, which used a very different approach to describe the gauge-invariant phase space of the theory. 

The Dirac bracket obtained from the particle gauge fixing condition also serves as a model for open universes, \ie spacetimes that have a boundary at spatial infinity  and on which one imposes the boundary condition that the associated metric is effectively conical. While we obtained the conical structure of the spacetime via a gauge fixing procedure, the Poisson structure of Theorem \ref{thm:particledirac} on the constraint surface defined by \eqref{mupres} can be regarded independently of its origin. In this case, the  mass and spin variables of the gauge-fixed particles that arise in this bracket  play the role of parameters that describe the orientation of the  cone. It would be interesting to see if and how this Poisson structure is related to the results in \cite{Meusburger:2005in} which defines a model for  open universes constructed by coupling non-standard punctures to Chern-Simons theory.

One of our main motivations for deriving the Dirac bracket of (2+1)-gravity is the construction of  a  fully gauge-invariant quantum theory based on a quantisation of the reduced phase space with the Dirac bracket. As the Dirac bracket associated with our gauge fixing conditions
can be formulated in terms of vector fields and functions on several copies of the three-dimensional Lorentz group $PSL(2,\RR)$, the gauge-invariant phase space should be directly amenable to quantisation. This would provide a fully gauge-invariant quantum theory of gravity which includes an observer and which is described in terms of physically meaningful quantities.
In particular, it  would be interesting to see what role the geometrical restrictions on the particle's worldlines play in the quantum theory, and if there are consistency conditions that require inclusion of this case.

%%%%%%%%%%%%%%%%%%%%%%%%%%%%%%%%%%%

\section*{Acknowledgements}

The research of the authors is funded by the German Research Foundation (DFG)
through the Emmy Noether fellowship ME 3425/1-1.  Both authors are also members
of the Collaborative Research Center 676 ``Particles, Strings and the Early
Universe''.  We are grateful to Winston Fairbairn for helpful suggestions and
comments on a draft of this article and thank Bernd Schroers and Philippe Roche
for remarks and discussions.  Some of our computations were double-checked with
the \textsc{xAct} tensor calculus package \cite{Martin-Garcia:2008aa}.

%%%%%%%%%%%%%%%%%%%%%%%%%%%%%%%%%%%

%%%%%%%%%%%%%%%%%%%%%%%%%%%%%%%%%%%

\bibliography{gaugefix}{}
\bibliographystyle{custom}

\end{document}